\documentstyle[12pt,aasms4]{article}

\received{5 February 1998}
\accepted{28 July 1998}
\slugcomment{To appear in the Astronomical Journal}

\begin{document}

\title{The Recent Star Formation in Sextans A}

\author{Schuyler D.~Van Dyk\altaffilmark{1}} 
\affil{IPAC/Caltech, Mail Code 100-22, Pasadena, CA  91125}
\authoremail{vandyk@ipac.caltech.edu}

\author{Daniel Puche\altaffilmark{1}}
\affil{Tellabs TG, 3403 Griffith, St-Laurent, Quebec H4T 1W5, Canada}
\authoremail{Daniel.Puche@tellabs.com}

\and

\author{Tony Wong\altaffilmark{1}}
\affil{Astronomy Department, University of California, Berkeley, CA  94720-3411}
\authoremail{twong@astro.berkeley.edu}
\altaffiltext{1}{Visiting Astronomer, Kitt Peak National Observatory, National
Optical Astronomy Observatories, operated by AURA, Inc., under contract with the
National Science Foundation.}

\begin{abstract}
We investigate the relationship between the spatial distributions of stellar populations
and of neutral and ionized gas in the Local Group dwarf irregular galaxy Sextans A.  This
galaxy is currently experiencing a burst of localized star formation, the trigger of which 
is unknown.  We have resolved various populations of stars via deep $UBV(RI)_{\rm C}$ imaging 
over an area with diameter $\sim 5{\farcm}3$.  We have compared our photometry with theoretical
isochrones appropriate for Sextans A, in order to determine the ages of these populations.  We 
have mapped out the history of star formation, most accurately for times $\lesssim$100 Myr.  
We find that star formation in Sextans A is correlated both in time and space, especially for 
the most recent ($\lesssim$12 Myr) times.  The youngest stars in the galaxy are forming 
primarily along the inner edge of the large H~I shell.  Somewhat older populations, 
$\lesssim$50 Myr, are found inward of the youngest stars.  Progressively older star formation,
from $\sim$50--100 Myr, appears to have some spatially coherent structure and is more centrally
concentrated.  The oldest stars we can accurately sample appear to have approximately a uniform
spatial distribution, which extends beyond a surface brightness of $\mu_B \simeq 25.9$ mag 
arcsec$^{-2}$ (or, a radius $r \simeq 2{\farcm}3$).  Although other processes are also possible,
our data provides support for a mechanism of supernova-driven expansion of the neutral gas, 
resulting in cold gas pileup and compression along the H~I shell and sequential star formation
in recent times.  
\end{abstract}


\section{Introduction}

Dwarf galaxies, both irregular and spheroidal, are the most numerous galaxies in the 
Universe.  Dwarf irregular galaxies play a key role in our understanding of galactic 
evolution, especially with the discoveries of faint, blue galaxies in high-redshift 
imaging surveys, such as the {\it Hubble Deep Field}.  Since dwarf irregular galaxies
are simpler, less-evolved systems, it should be possible to observe short-lived 
evolutionary properties in nearby systems, which would be more easily destroyed in 
more massive, complex systems, such as grand design spiral galaxies. Knowledge of the
star formation history of dwarf irregular galaxies is of particular importance; Hunter
(1997) recently presented a comprehensive discussion of the topic.

Dwarf irregular galaxies show a larger range in star formation rates than do
spiral galaxies.  Studies of the stellar populations in dwarf irregulars (e.g., 
\cite{tos91}, \cite{gre93}, \cite{mar95}) show that most of these galaxies evolve with 
approximately constant star formation rates, either high or low, interrupted by 
variations of factors 2--3 in these rates over the past few Gyrs (Hunter 1997).  The 
question, then, is what governs and regulates star formation in dwarf irregulars.  
A vital clue to answering this question can be found in the relationship between the 
structure and kinematics of the gas and the history of star formation.  The gas can be
traced and analyzed by mapping the neutral, molecular, and ionized components, and 
the star formation history can best be traced by the distributions of resolved 
stars of various ages.

A program to study the interstellar medium (ISM) in dwarf irregular galaxies was 
undertaken by D.~Puche, D.~Westpfahl, and their collaborators, with the goal of 
determining if the neutral component of the ISM is distributed differently in low-mass
systems than in grand design spirals.  Puche et al.~(1992), Westpfahl \& Puche (1994),
and Puche \& Westpfahl (1994) found that, based on their high-resolution HI data for 
a number of dwarf galaxies, the neutral gas component shows an unprecedented amount 
of structure, in the form of shells or bubbles (\cite{hei79}, 1984; \cite{ten88}).
Low rotation velocities for dwarfs means less shear, and thus, longer lifetimes for
these large-scale shells.  Furthermore, the gas scale-height in dwarf galaxies must be
much larger than in high-mass spirals, resulting in a lower gas density and a higher 
efficiency for making large holes (\cite{hei79}).  Also important, of course, is the 
absence of spiral density waves (see Hunter, Elmegreen, \& Baker 1998).

Puche et al.~(1992), Puche \& Westpfahl (1994), and Westpfahl \& Puche (1994) 
found that these structures are expanding and argued that they surround hot, 
low-density bubbles, formed as the result of the combined effects of photoionization,
stellar winds, and multiple supernovae from sequential formation of massive stars 
(\cite{cas80}).  Supernovae would likely dominate all other sources of energy input,
although winds from massive stars also substantially contribute (e.g., \cite{lei92}).
The models of Chevalier (1974) and Koo \& McKee (1992a,b) predict that about 100 
supernovae exploding in pre-existing wind-blown bubbles would provide about $10^{53}$ 
erg to the ISM and can produce shells with radii of hundreds of parsecs, expansion 
velocities of $\sim$10--20 km s$^{-1}$, and lifetimes of $\sim 10^8$ yr,
consistent with those observed.  

If supernovae are producing expanding bubbles, Puche \& Westpfahl have argued,
the current star formation in these galaxies should be concentrated near the 
high-density regions of the expanding H~I shell at the edge of the inner bubble.
This is consistent with theories of sequential star formation (e.g., 
\cite{ger78}; \cite{ger80}), in which the compression associated with the shell 
triggers the star formation process.  If the supernova-driven scenario holds, 
then we would expect star formation episodes to be not only correlated in age, 
but also in spatial distribution, along the H~I shell structures.

Limited, low-resolution optical data tend to suggest that this scenario is 
possibly taking place:  H$\alpha$ appears concentrated on the edges of the nearly 
circular H~I shells in Holmberg II (\cite{puc92}; \cite{hod94b}), Holmberg I 
(\cite{wes94}), and IC 2574 (\cite{mar94}).  Another, more local, example where this
also may be occurring is Constellation III and the supershell LMC 4 in the LMC 
(\cite{dop85}, \cite{gre98}; although see \cite{rei87}, \cite{bra97}, \cite{ols97}).
Additionally, Puche et al.~(1992) claim that broad-band imaging shows the 
possible presence of stars or clusters at the centers of some of the largest 
shells in Holmberg II, which could be the populations, associated with the
supernovae progenitors, which remain long after the explosions.  

One of the galaxies Puche \& Westpfahl (1994) sampled was Sextans A (DDO 75, 
UGC 205, A1008$-$04), a low surface brightness dwarf irregular galaxy with 
angular diameter of only $D_{25} \sim 4{\farcm}4$ (\cite{abl71}; \cite{tul88}) and 
an oddly square, or, more appropriately, diamond shape.  A large number of resolved 
stars can be seen on a number of published ground-based (and space-based) images of
the galaxy.  The galaxy is seen nearly face-on (inclination $i \sim 36\arcdeg$,
\cite{ski88}; to $45\arcdeg$, \cite{gra98}, in preparation).  A recent estimate of 
the distance to Sextans A by Sakai, Madore, \& Freedman (1996; SMF, hereafter), 
based on the tip of the red giant branch (TRGB) brightness, is 1.42 Mpc 
($m-M=25.74$), placing it near the edge of the Local Group.  At this distance, the
angular diameter corresponds to a linear diameter of only $\sim$1.8 kpc for the 
galaxy.  The galaxy has $B_T=11.93$ and $(B-V)_T=0.30$ (somewhat bluer than the mean
for Magellanic irregulars; \cite{abl71}); at the TRGB distance modulus, 
$M_{B_T}=-13.81$.  The morphology of Sextans A in H~I had been initially studied by 
Skillman et al.~(1988).  Skillman et al.~estimate a gas mass of 
$M \sim 5.8 \times 10^7\ M_{\sun}$ and a total mass of 
$M \gtrsim 1.6 \times 10^8\ M_{\sun}$.

From higher resolution and greater sensitivity maps, Puche \& Westpfahl (1994)
found that Sextans A looks similar in H~I morphology to other small dwarf 
irregulars, such as Holmberg~I and M81dwA (\cite{wes94}), in that one H~I hole
appears to dominate the overall structure, although Sextans A is a slightly 
faster rotator than these other small dwarfs ($v_{\rm rot\ max} \simeq 30$ km 
s$^{-1}$ for Sextans A, compared to $\sim$15 km s$^{-1}$ for Ho~I and $\sim$5
km s$^{-1}$ for M81dwA; \cite{puc94}), consistent with the central large cavity
appearing to have undergone some shearing and twisting.

This galaxy has also been the object of previous intense optical study, both from 
the ground and, recently, from space.  Its resolved stellar populations have been 
observed in a number of different bands.  Sandage \& Carlson (1982, 1985; hereafter, 
SC82 and SC85, respectively; SC85 revises the photometry presented in SC82) 
published an early photometric study, consisting of photographic
$BV$; Hoessel, Schommer, \& Danielson (1983; hereafter, HSD) obtained CCD 
Gunn-Thuan $gri$ photometry; Walker (1987; hereafter, W87) published CCD $BV$ 
photometry of bright stars; Aparicio et al.~(1987; hereafter, A87) presented deep 
CCD $UBV$ photometry; and SMF recently analyzed deep CCD $BVI$ photometry.  Based 
on their deep $UBV$ imaging data, A87 presented a history of recent star formation 
in Sextans A, but they considered the star formation over only quite broad regions 
of the galaxy.  SMF, although their $BVI$ imaging was quite deep, did not
address in depth the star formation in Sextans A, dealing primarily with their
distance estimate.

Recently, Dohm-Palmer et al.~(1997a,b) presented reports on photometric
results based on deep {\sl Hubble Space Telescope\/} (HST) imaging.  Dohm-Palmer
et al. (1997b) provide a detailed description of the star formation history in 
Sextans A over the last 1 Gyr, and illustrated the correlation of the
history both in time and in space.  Their results are unhampered by the problems of 
crowding and blending which beset ground-based studies, and the number of
stars that they can statistically analyze is truly formidable.

Based on their results, Dohm-Palmer et al.~(1997b) conclude that from 100 -- 600 Myr 
ago, the star formation rate ($\sim$2000 $M_{\sun}$ Myr$^{-1}$ kpc$^{-2}$) was about 
six times larger than the rate ($\sim$310 $M_{\sun}$ Myr$^{-1}$ kpc$^{-2}$) averaged
over the lifetime of the galaxy, and that since about 80--100 Myr ago, the rate has 
become remarkably higher (by a factor $\sim$20, at $\sim$5000 $M_{\sun}$ Myr$^{-1}$ 
kpc$^{-2}$) and more densely concentrated than in the past, suggesting that some 
unusual event has occurred in the past 100 Myr to trigger the recent star formation.
(Their movie of star formation in \cite{doh98} seems to indicate that the trigger 
might have begun as early as $\sim$350 Myr.)

Dohm-Palmer et al.~speculate on possible scenarios for this mysterious event, including
a tidal interaction with neighboring Sextans B; however, they point out that these two
galaxies may be separated by nearly 300 kpc, a distance over which the dynamical effects
would be minimal.  The trigger could also be the recent onset of a bar potential, which 
Skillman et al.~(1988) suggest possibly exists in Sextans A, based on their data; although
Puche disagrees, Westpfahl (private communication) states that the gas kinematics based on
their recent H~I data could also indicate the presence of a bar.  Graham \& Westpfahl (1998)
suggest a possible merger or accretion event.  Finally, the supernova-induced star formation
mechanism could be the trigger for recent star formation.  Combinations of one or more of 
these scenarios is, of course, also possible.

In this paper, we present deep $UBV(RI)_{\rm C}$ imaging of Sextans A, in which a large number
of stars are resolved.  Through our ground-based five-band imaging we have sampled 
nearly the entire main body of the optical galaxy, and therefore we can more completely 
describe the temporal and spatial history of star formation over a large area, particularly 
for the last 100 Myr, and (with far less certainty) back to about 1 Gyr.  Furthermore, we 
examine the correlation of the recent star formation history with both the neutral H~I and 
the ionized gas components.  

\section{The Data}

\subsection{Observations and Reductions}

The observations were conducted at the Kitt Peak National Observatory (KPNO)
2.1-m telescope in $UBV(RI)_{\rm C}$ on 1994 Feb 11 UT, as part of 
a program to image several dwarf irregular galaxies showing significant 
structure in H~I.  The KPNO ``T1KA'' $1024^2$ CCD was used in direct 
imaging mode at $f/7.5$; the pixel scale was $0{\farcs}31$, resulting in a 
field size of $5{\farcm}2$.  An area of Sextans A with diameter $\sim 5{\farcm}3$ 
could therefore be imaged on the chip.  All images exhibited large numbers
of resolved stars in the galaxy.  Total exposures in each band were divided into three
separate exposures, dithering the telescope slightly between exposures
to reduce pixel-to-pixel and position-dependent effects on stellar profiles and
magnitudes.  Total exposure times were 1800 sec in each of the five filters.
Unfortunately, although the seeing was good 
($\sim 1{\farcs}1$), the night was not photometric (nor was the entire run).
Figure 1 shows a three-color composite image of the galaxy.

Observations were subsequently made on 1995 April 24 UT using the Lick 
Observatory Nickel 1.0-m telescope in $BV(RI)_{\rm C}$, under photometric 
conditions, for the purpose of calibration.  The Nickel telescope imaging system,
consisting of an Orbit $2048^2$ CCD (binned $2 \times 2$), with pixel scale 
$0{\farcs}37$ and field size $6{\farcm}3$, however, has essentially no response 
in $U$.  Fortunately, a service observing campaign (1996 April 1--5 UT) was 
undertaken during bright time at the KPNO 0.9-m telescope by Mr.~R.~Young Shuping,
to provide calibration observations for previous KPNO imaging programs 
conducted under non-photometric conditions.  Shuping obtained 
$UBV(RI)_{\rm C}$ images for us.  Shorter (600--900 sec) exposures of Sextans A
were made during both sets of calibration observations.

Additionally, we obtained H$\alpha$+[N II] and $R$-band images of Sextans A at 
the old KPNO ``No.~1'' 0.9-m telescope in 1989 May 31 UT.  These observations
employed the old ``RCA3'' $508 \times 312$ chip, with pixel scale $0{\farcs}86$
and field size $7{\farcm}3 \times 4{\farcm}5$.  Three individual images through
a narrow-band filter with central wavelength 6563 \AA\ (bandpass 75 \AA) were made.
The $R$-band image taken during that observing run had an exposure time
of 300 sec.  

The usual bias and flatfield images were obtained during all observing runs.
All image processing was performed using IRAF\footnote{IRAF (Image Reduction
and Analysis Facility) is distributed by the National Optical Astronomy
Observatories, which are operated by the Association of Universities for
Research in Astronomy, Inc., under cooperative agreement with the National
Science Foundation.}.  
As mentioned above, the separate 2.1-m exposures of Sextans A were 
combined into a single total image, using cosmic-ray rejection,
for each filter.  Additionally, the three H$\alpha$+[N II] images were
combined into a single 1500-sec image.

A continuum-subtracted H$\alpha$+[N II] image was produced using the combined
narrow-band ``on-line'' observation and the broad-band $R$ observation, where
$R$ was employed as the ``off-line'' filter, assuming that H$\alpha$+[N II]
are the only emission lines dominating the $R$ bandpass (see \cite{wal90}).
The scale factor for the subtraction was derived from the intrumental
photometry of the field stars in common between the two sets of images.
The subtracted image was not flux calibrated.

\subsection{Photometry Methods}

The DAOPHOT and ALLSTAR (\cite{ste87}; \cite{ste90}) routines in IRAF were used to 
derive the magnitudes of the resolved stars on all the $UBV(RI)_{\rm C}$ images 
made of Sextans A (including those made with the KPNO 2.1-m, the Lick 1-m, and
the KPNO 0.9-m telescopes).
These routines perform multiple fittings of a semi-empirical point spread 
function (PSF) to the individual stellar profiles.  The PSF was established by
assuming a Moffat analytical function, plus a table of residuals, using a number
($\sim$30) of well-shaped stars on each image.  Figure 2 shows the residuals of the
fitting, $\sigma$, given by ALLSTAR for all stars in each band on the images
made at the KPNO 2.1-m telescope (after elimination of probable foreground stars,
obviously blended stars, and erroneous detections; see \S 3 below).  
These $\sigma$ values are representative of the internal errors in the photometry 
presented in this paper.

Our Lick 1-m and Shuping's KPNO 0.9-m telescope observations not only consisted of
images of Sextans A, but also images of a number of standard star fields, 
spanning a wide range of colors, from Landolt (1992).  The standard star instrumental
magnitudes were determined through large (radius $\sim$7\arcsec) synthetic apertures, 
which significantly included the wings of the stellar profiles, and sky annuli with radii
larger than the apertures.  The photometric 
transformations to the absolute system were computed using these instrumental magnitudes.

The ALLSTAR PSF-fitting magnitudes for stars in Sextans A imaged during the two sets of
calibration observations were corrected to synthetic aperture 
magnitudes, using the aperture-curves-of-growth method (Stetson 1990).  However, this method
was only reliable for the $\sim$15 brightest stars (or fewer) on these shorter-exposure 
images of the galaxy in each band.  The ALLSTAR magnitudes from the KPNO 2.1-m
images were, of course, not aperture-corrected, since these observations were not 
photometric, but had to be ``bootstrapped'' to the absolute system using the observations
of Sextans A during the calibration runs.

\subsection{Transformation to the Standard Johnson-Cousins System}

For the Lick calibration run, an adequate number (30) of standard stars were observed 
over a large range in airmass.  
The aperture magnitudes of the standard stars were then used to determine the 
atmospheric extinction in each band and the transformation to Johnson-Cousins 
magnitudes.  The transformation equations, derived using the PHOTCAL tasks in
IRAF, for the Lick Nickel 1-m run are

\begin{equation}
b=B + 5.234[{\pm}0.051] + 0.315[\pm{0.032}] X - 0.057[\pm{0.070}] (B-V) - 0.008[\pm{0.040}] (B-V) X
\end{equation}

\begin{equation}
v=V + 4.926[{\pm}0.019] + 0.196[{\pm}0.012] X + 0.044[{\pm}0.025] (B-V) - 0.005[{\pm}0.016] (B-V) X
\end{equation}

\begin{equation}
r=R_{\rm C} + 4.858[{\pm}0.029] + 0.134[{\pm}0.019] X + 0.035[{\pm}0.070] (V-R_{\rm C}) + 0.001[{\pm}0.042] (V-R_{\rm C}) X
\end{equation}

\begin{equation}
i=I_{\rm C} + 4.908[{\pm}0.053] + 0.112[{\pm}0.029] X + 0.069[{\pm}0.127] (R-I)_{\rm C} - 0.049[{\pm}0.070](R-I)_{\rm C} X
\end{equation}

\noindent where the lower-case letters represent the aperture (instrumental)
magnitudes and the upper-case letters represent the true Johnson-Cousins 
magnitudes, and $X$ is the airmass of the observation.  The formal errors in the 
coefficients are also given.

Although standard stars over a large range of color were observed during Shuping's 
photometric night on the KPNO 0.9-m, a smaller number (17) of stars were observed without
adequately sampling over airmass.  As a result we had difficulty achieving a reasonable 
solution for the photometric coefficients, while allowing them all to be free variables.  
We therefore chose to fix the airmass corrections to the nominal values appropriate for
KPNO, as given in the ``Direct Imaging Manual for KPNO'' (P.~Massey et al.~1997).  The 
transformation equations are then

\begin{equation}
u=U + 6.972[{\pm}0.030] + 0.500 X - 0.076[{\pm}0.248] (U-B) - 0.055[{\pm}0.167] (U-B) X
\end{equation}

\begin{equation}
b=B + 4.902[{\pm}0.007] + 0.250 X - 0.019[{\pm}0.039] (B-V) - 0.064[{\pm}0.028] (B-V) X
\end{equation}

\begin{equation}
v=V + 4.665[{\pm}0.004] + 0.150 X + 0.134[{\pm}0.020] (B-V) - 0.095[{\pm}0.013] (B-V) X
\end{equation}

\begin{equation}
r=R_{\rm C} + 4.579[{\pm}0.007] + 0.100 X + 0.115[{\pm}0.064] (V-R_{\rm C}) - 0.124[{\pm}0.047] (V-R_{\rm C}) X
\end{equation}

\begin{equation}
i=I_{\rm C} + 5.353[{\pm}0.007] + 0.070 X + 0.005[{\pm}0.058] (R-I)_{\rm C} - 0.082[{\pm}0.044] (R-I)_{\rm C} X
\end{equation}

\noindent where the symbols are as above.  The formal are also given for those coefficients
treated as free parameters.

The aperture-corrected ALLSTAR magnitudes for the stars resolved on the images 
of Sextans A made during the Lick 1-m and KPNO 0.9-m runs were then transformed 
to the Johnson-Cousins system using the PHOTCAL routines within IRAF.  We
compared the final photometry results from the Lick 1-m and KPNO 0.9-m 
observations, for stars in Sextans A in common between the two sets (we were limited to a 
comparison for only the brightest stars, i.e., those with $V \lesssim 17$)
and found that the $BVRI$ magnitudes and colors agreed between the two sets
to the following degree: $-0.003 \pm 0.024$ [$V$], $0.013 \pm 0.033$ [$B-V$], 
$-0.019 \pm 0.014$ [$V-R$], and $0.000 \pm 0.025$ [$R-I$] (the sense of the
sign is ``KPNO 0.9-m$-$Lick 1-m'').  One can see that the agreement is quite good.
We therefore used the photometry from the Lick run to transform the ALLSTAR magnitudes
obtained from the deep KPNO 2.1-m imaging in $BV(RI)_{\rm C}$.

We had no choice for $U$; the agreement between the $BV(RI)_{\rm C}$ calibration obtained
from the Lick 1-m and KPNO 0.9-m imaging provided us with confidence that
the $U$ calibration done on the KPNO 0.9-m was reasonable.  We therefore
transformed the $U$ photometry from the KPNO 2.1-m images using the photometry
from the KPNO 0.9-m image.

\subsection{Comparison with Previous Photometry}

In any study of this kind, it is always appropriate to compare the new photometry
with older published photometry when available.  Fortunately, as mentioned above,
Sextans A has been the subject of a good number of recent optical studies.  Here,
we can compare subsets of our photometry with the results from SC82, SC85, HSD 
(their $gri$ magnitudes were transformed into Johnson $BVRI$ magnitudes, via the 
transformations in Hoessel \& Melnick 1980 and Wade et al.~1979), W87, A87, and SMF.
(Dohm-Palmer et al.~1997a,b present $BVI$ photometry using HST, but they do not
provide their data in tabular form, so we do not make a comparison with
their photometry.)  

The stars which have been most used as the basis of comparison in
all these previous studies are the bright stars listed in SC82.  
In Table 1 we list the mean differences of our measurements for the various
colors for the SC82 stars with the measurements from these other studies.
In Figure 3 we show a comparison of our magnitudes and colors for the
SC82 stars with those obtained by the other studies.

One can see that our $V$ magnitudes are in very good agreement with those of SMF, 
A87, and W87, but differ substantially from SC85 and HSD, the former likely due to
the zero-point problems associated with the photographic photometry, and the latter 
likely due to the uncertainty in the transformation from the Gunn-Thuan system to the 
Johnson system, as noted by SMF.  Our $B-V$ colors agree very well with those in W87,
but less so with those from the other studies.  Our $U-B$ colors are in reasonably 
good agreement with those from A87, but in the case for both $U-B$ and $B-V$, 
we point out the large amount of scatter for even the brightest stars studied in
A87.  

One can see very good agreement of our $V-I$ colors with those from SMF, with
an offset of $\sim 0.05$ between our colors and theirs (although the rms in the comparison
is roughly equivalent to this possible offset; additionally, we notice a similar offset
between the $I$ magnitudes from SMF and those from Dohm Palmer et al.~1997a in the 
latter's Figure 7 for $I \lesssim 20.3$).  Finally, we compare our $R-I$ (Cousins system) 
colors with the $R-I$ (Johnson system) colors from HSD.  The offset of $\sim 0.1$ 
%may be due both to the uncertainty in the transformation from the Gunn-Thuan system to the
%Johnson system and to the complex nature of the transformation between the Johnson 
%system and the Cousins system (\cite{bes83}), but it 
agrees with the zero-point of 0.12 in the transformation from the Johnson to the 
Kron-Cousins system derived using asymptotic giant branch stars by Costa \& Frogel 
(1996).

In all, we believe the agreement of our photometry with previous photometry
to be quite satisfactory.

\subsection{Transformation to Absolute Coordinates}

Absolute star positions 
were derived by converting the pixel centroids of the stars on our KPNO 2.1-m 
images from ALLSTAR into absolute coordinates using the Guide Star Astrometric 
Program (GASP) software in 
STSDAS/IRAF and the Digitized Sky Survey\footnote{The digitizations of Schmidt 
Sky Survey plates have been made by the Space Telescope Science Institute (STScI)
under contract NAS5-26555 with the National Aeronautics and Space Administration
and US Government grant NAG W-2166.  The images of these surveys are based on 
photographic data obtained using the Oschin Schmidt Telescope on Palomar Mountain
and the UK Schmidt Telescope for the Palomar Observatory
Sky Atlas (POSS-I), made by the California Institute of Technology with grants
from the National Geographic Society.}
image of the Sextans A field obtained from the
Canadian Astronomy Data Center (CADC), from which absolute coordinates for stars
on the Palomar Observatory Sky Survey can be measured.  The accuracy
obtained in these absolute positions by this technique ($\sim 1$--$2\arcsec$)
is adequate for our
comparison with the H~I radio map, given the relatively lower resolution of the
radio data (see \S 4).

\section{Analysis of the Photometry}

We present here a brief description of our photometric results.  We have 
removed obvious foreground stars from our star list, e.g., the notoriously bright
red foreground star in the northeast quadrant of Sextans A, and 
those stars that A87 and SC82 previously identified as foreground stars (e.g.,
those SC82 measured as proper motion stars).  These
are essentially all stars with $V \lesssim 17.5$.

We can estimate how many foreground stars fainter than this $V$ should be in
our field, from the Ratnatunga \& Bahcall (1985) model for the galactic 
coordinates of Sextans A (we interpolate the model appropriate for Sextans A
from the nearby fields including the galaxies Pal 3 and Leo I).  In Table 2
we list, for our 27 arcmin$^2$ field, the predicted number of stars in three
color bins and four magnitude bins, along with the estimated number of actual
stars found within these bins.
Of course, without identification spectra for individual stars, we have no means
of eliminating possible foreground interlopers from our star list, but the 
comparison in Table 2 should be kept in mind throughout the paper.

Finally, we have also used our color-color diagrams (\S 3.2) to isolate and remove
stars from our list which clearly have grossly discordant colors, indicating possible
blends, unresolved nebulosity, or other photometric errors.  We have eliminated 
erroneous detections along the two bad rows on the combined images and within the 
saturated image of the aforementioned bright red foreground star; we have also removed
stars that are too close to the edge of the CCD in any band to provide reliable photometry.

We arrive at 2525 stars detected in $V$ and at least one other band.
For space reasons,
we do not list the magnitudes and colors for all the stars here.  
An electronic tabulation of magnitudes, colors, and absolute positions of the stars 
can be obtained courtesy of the CDS.  

\subsection{Luminosity Functions}

We can derive a luminosity function for the stars in the galaxy and compare it to 
those for other dwarf irregulars.  In Figure 4 we show the global differential 
luminosity function for all detected stars in each of the bands.  Crowding and 
blending of stars in our images limit the depth and accuracy of the photometry we 
have obtained.  This can be represented by the completeness of our photometry.
One can see that incompleteness significantly affects the photometry in each of the 
bands at a given magnitude.  Our photometry is most complete in the 
$V$-band, next most complete in $R$, then $I$, $B$, and, finally, $U$.  The 
photometry is seriously affected by incompleteness at $V \sim 23$, $R \sim 22.5$, 
$I \sim 22$, $B \sim 21.7$, and $U \sim 20.7$.

We also derive the luminosity function for only the main sequence stars.
Following the recommendation by Freedman (1985), we separate these stars from the 
evolved ``blue loop'' supergiants (which are indistinguishable from the main sequence
stars in the blue plume on the $(V,B-V)$ CMD in Figure 6) by selecting those stars 
with $U-V \lesssim -1.05$ (we use a similar criterion in $U-B$ in \S 3.3 to distinguish
the hot main sequence stars).
In Figure 4 we show the differential LF for only those 241 stars 
({\it filled circles}).  A least-squares fit to the data for the main sequence 
stars brighter than $V=22$ (fainter than which incompleteness clearly
affects our statistics) results in a slope, $d\log N / dV = 0.48$.

This is consistent with the slope for other dwarf irregulars, such as WLM, Sextans
B, and NGC 6822 (see, e.g., \cite{tos91}; \cite{gre93}; \cite{mar95}), and implies 
that Sextans A has a star formation rate and initial mass function that is not unlike
other dwarf irregulars that have been studied so far.  As pointed out by Hunter \& 
Plummer (1996), Sextans A currently has a star formation rate ($\sim 0.006 M_{\sun}$ 
yr$^{-1}$), when normalized to its linear scale, which is typical of other dwarf 
irregulars (see also \cite{hun86}), being neither very low nor excessively high.

Interestingly enough, similar slopes are found for the luminosity functions for all 
stars in each of the bands, for which the photometry is complete (0.44, for all
stars in $V$; 0.44, for $B$; 0.53, for $U$; 0.48, for $R$, and 0.53, for $I$).  This
implies that the mass function averaged over more than
1 Gyr has been relatively constant in Sextans A.  We have also isolated just those
stars along the putative bar in the galaxy; we find that the slope of the luminosity
function for these stars in $V$ is 0.51, similar to the slopes in $V$ and other
bands for all stars in Sextans A.

\subsection{Color-Color Diagrams}

In Figure 5 we present the color-color diagrams for Sextans A.  The 
{\it solid line\/} on each diagram shows the location of the (unreddened) main 
sequence, giant and supergiant branches, based on theoretical models (see
\S 3.4).  We have indicated on these color-color diagrams the direction of the 
reddening vector.  We also show representative uncertainties in the observed colors.
On the ($U-B,B-V$) diagram we can see two clumps of stars, 
the bluer main sequence stars and the somewhat redder blue supergiants.  On the 
($B-V,V-I$) diagram, we again see a blue clump of stars, which is a mixture of main 
sequence and supergiants, and a smaller clumping of red supergiant stars.
The general agreement between the model tracks and the observed color distributions 
indicates that the amount of reddening to and within the galaxy must be 
generally small (see \S 3.3).
But, some reddening is clearly necessary, based on these diagrams, on the amount of 
Galactic foreground reddening, and on the results of other investigators from
colors of individual stars (e.g., SMF).  The reddening also appears somewhat variable. 
 
\subsection{Color-Magnitude Diagrams}

The most direct information on the stellar populations and relative star 
formation
histories in galaxies can be obtained from the analysis of color-magnitude
diagrams (CMDs) based on deep photometry of the resolved populations.  In
this paper we have observed at five optical bands:  our bluest colors can
very adequately trace out the youngest, hottest stars in the galaxy,
while, as Aparicio \& Gallart (1995) have demonstrated in their study
of the Pegasus dwarf galaxy, photometry in red colors is also necessary,
to derive information about stars of all ages.

In Figure 6 we present the CMDs for Sextans A in the various colors based on 
our photometry.  What is immediately evident from examining all of the
CMDs, as has been found in previous studies of Sextans A, is the rich and 
various array of populations of different ages, from main sequence stars to 
red supergiants to asymptotic giant branch (AGB) stars.
Such populations have been identified in this and other dwarf galaxies from
the ground, but never before with this color baseline.

The practical detection limit for our photometry is $V \lesssim 23$.
In our deep ground-based images, a diffuse background of faint stars in the galaxy
is also discernible below this limit.

On the $(V,B-V)$ and $(V,V-R)$ CMDs one can clearly see the blue sequence, 
or ``plume,'' of stars, with $B-V \lesssim 0.2$ ($V-R \lesssim 0.3$).  The 
greatest limitation to separating out the various populations in Sextans A 
based on our imaging is, of course, crowding and blending.  Despite this 
limitation in resolution, however, our $(V,U-B)$ CMD clearly shows that this blue
plume separates into the main sequence (with $U-B \lesssim -0.7$), down to about
spectral type B1 ($V \sim 22$), and the evolved ``blue loop'' supergiant stars,
with $-0.7 \lesssim U-B \lesssim 0.3$, which are burning helium (He) at their cores,
although some small contamination exists from main-sequence turn-off stars.
(This blue loop is not as evident in our $[V,B-V]$ and $[V,V-R]$ CMDs, due in part
to the above-mentioned crowding and blending of main sequence and evolved stars on 
the CCD images, but also to the fact that $U-B$ affords a much better temperature 
resolution for luminous blue stars and therefore assists in overcoming the degeneracy
in the blue plume.)

{\it We therefore have been able to detect the brighter\/} ($V \lesssim 22$)
{\it blue loop population based on our ground-based data, even in a galaxy as distant
as Sextans A through deep $UBV$ imaging.}

On the blue CMDs one notices the very bright star with $V=17.51$, $U-B=-0.03$, 
and $B-V=0.14$; A87 point to it likely being a supergiant member of 
Sextans A, primarily due to its location in the bright blue clustering of stars 
in the southeast.  The PSF for this object does appear stellar, so it is not likely
to be a blend.  For this star $M_V \simeq -8.3$ for a distance modulus of 25.8
(see \S 3.4).  It appears therefore to be $\sim$2 mag brighter than the blue straggler 
binary stars found in the young SMC cluster NGC 330 (Grebel, Roberts, \& Brandner 1996).
It is more likely to be a luminous blue variable star, similar to the one found by Drissen,
Roy, \& Robert (1997) in NGC 2366.  This star clearly deserves more attention.

One can also see on the $(V,B-V)$ CMD, and, more so, with 
the addition of the $R$ band in the $(V,V-R)$ CMD, that the number of blue 
loop stars is relatively larger than the number of red supergiants, in a ``red 
plume'' of stars with $B-V \gtrsim 0.5$ ($V-R \gtrsim 0.6$) and $V \lesssim 22$.  
More blue supergiants than red supergiants may exist in this galaxy.  This is 
consistent with the fact that, at low metallicity, stars spend more time as blue 
supergiants than red supergiants during their He core-burning phase (\cite{ber94}).

On the $(I,R-I)$ and $(I,V-I)$ CMDs one, again, can see the conspicuous red 
``plume'' of stars, at $0.8 \lesssim V-I \lesssim 1.6$
($0.3 \lesssim R-I \lesssim 0.8$) and $I \lesssim 22.3$, with two or three 
conspicuous clumps of red supergiants, as noted by SMF, at $I \sim 21.2$, 20.6, 
and 19.8.  Also visible is the broad clump of red stars, at 
$0.8 \lesssim V-I \lesssim 1.8$ ($0.3 \lesssim R-I \lesssim 0.8$),
$I \sim 22$, which is the top of what Aparicio \& Gallart (1995) refer to as
the ``red tangle'' of RGB stars, old and intermediate-age AGB stars, and 
intermediate-age blue-loop stars.  The TRGB, although
likely contaminated by intermediate-age AGB stars, is at $I \sim 21.8$ and 
$V-I \sim 1.3$ (SMF).  

The stars fainter than $I \sim 19$, with $V-I \lesssim 2$ ($R-I \lesssim 0.9$) 
are what Aparicio \& Gallart (1995) refer to as the ``red tails'' of 
intermediate-age and old AGB stars.  Aparicio \& Gallart point out that the 
``length'' of the ``tail'' (i.e., how red this feature extends on the CMD) is a 
function of metallicity; the longer the tail, the greater the metallicity.
The observed tails for Sextans A do not extend as red in color as those on the 
CMDs for, e.g., the Pegasus dwarf galaxy (\cite{apa95}), consistent with the 
metallicity difference between Sextans A and Pegasus (Skillman, Bomans, \&
Kobulnicky 1997; Aparicio, Gallart, \& Bertelli 1997).

Of particular curiosity are the very bright, red stars seen on the red 
CMDs.  Although they appear to be especially, and possibly anomalously,
bright, to be members of Sextans A, we have no particular evidence to the
contrary, e.g., they are not among the proper motion stars in SC82.  All but one
of the stars is in or near a region of recent star formation.  Short of
having spectra for these stars, we consider them to be members of the galaxy.

We also show on all CMDs, highlighted with {\it open circles}, the magnitudes
and colors of the known Cepheids from Piotto, Capaccioli, \& Pellegrini (1994).

\subsection{Stellar Population Ages}

On the CMDs in Figure 6, we have compared our photometry with a series of 
theoretical isochrones calculated by the Padova group (see \cite{ber94} and 
references therein), which we use to estimate the ages of the stellar 
populations in Sextans A.  We have chosen those isochrones computed with a 
metallicity $Z=0.001$, which is consistent with that inferred from the oxygen 
abundance for this galaxy (\cite{ski89}).  We do not extensively sample the 
oldest populations, where consideration of significantly lower metallicity 
would be important.  We have found that isochrones of higher metallicity do not 
match the positions of the stars in color and magnitude nearly as well as those 
of the chosen metallicity.

Before further comparison with the isochrones is made, in order to estimate
ages for the various populations, it is first necessary to choose a distance to 
the galaxy and to consider the amount of reddening that the observed stars are 
experiencing.  We therefore adjust the isochrones on the CMDs accordingly.

First, we have adopted a distance modulus to Sextans A which is a compromise 
between the modulus derived from the TRGB and Cepheid brightness methods, given by
SMF, i.e., $m-M \simeq 25.8$ mag (which corresponds to a distance of 1.44 Mpc).
Second, we have assumed a reddening for the galaxy of $E(B-V) = 0.05$.  The
amount of Galactic reddening is $E(B-V) = 0.02$ (Burstein \& Heiles 1984), but,
as SMF point out, $E(B-V) \simeq 0.05$ is likely appropriate for the main body
of the galaxy, where most of the young stars and gas are found (SMF find a 
higher $E[B-V]$ for the older red giant population in the galaxy).  So,
we have assumed this value, which, assuming a normal reddening law, i.e.,
$A_V = 3.1 E(B-V)$, corresponds to $A_V \simeq 0.16$.  Note that this is 
somewhat larger than the reddening assumed by Dohm-Palmer et al.~(1997a,b), 
which is essentially only due to the Galactic component.  Based on our
reddening assumption, 
we follow Cardelli, Clayton, \& Mathis (1989), in deriving the extinction for
the other bands, and assume that $E(U-B) \simeq 0.7E(B-V)$.

From examining all the CMDs it is evident, regardless of their exact ages, that
a number of stellar populations of various ages coexist in Sextans A, as has
been noted in previous studies (e.g., A87; Dohm-Palmer et al.~1997b).
The star formation does not seem to have been continuous, but instead appears 
to have occurred more in bursts or episodes.  We do not try to execute a thorough 
and exact analysis of the population ages through more sophisticated means, such as
synthetic CMDs (e.g., \cite{apa96}).  Based on our more cursory analysis with the 
theoretical isochrones, it can
be seen from the CMDs constructed from the $U$, $B$, and $V$ bands
that the blue plume of stars consists of several 
young populations, from very young (only $\sim$ a few Myr) to significantly
older, with ages $\lesssim 100$ Myr.  Clear gaps in stellar age within this
time range exist, as best seen in the redder populations (i.e., with
$B-V \gtrsim 0.4$).

The redder CMDs show three noticeable clumps of red supergiants, with ages
$\sim$40 Myr, $\sim$60 Myr, and $\sim$80 Myr.  The number of red supergiants
both older and younger than these ages appears to be significantly smaller.
These stars must have formed in three different bursts of recent star
formation.
The clump of RGB stars (red tangle), for which we can resolve only the top,
appears to correspond to an age of $\gtrsim 100$ Myr.  The
oldest populations 
we can resolve (the diffuse quantity of red stars with $V-I \gtrsim 1.5$,
comprising the various red AGB tails) appear to have ages from $\sim$0.2--0.3 
Gyr to several Gyr, although we can only probe these populations with accuracy
to ages 1--2 Gyr.  Our photometry is not deep or complete enough to resolve 
stars older than this.

From our CMDs we find that the Cepheids in Sextans A have ages
$\gtrsim$40 Myr and appear to be younger than $\sim$100 Myr.

\section{Spatial Distribution of Stellar Populations}

First, we examine the spatial distribution of the youngest population of stars
in Sextans A.  We consider stars with ages $\lesssim 50$ Myr.  Dohm-Palmer et 
al.~(1997b; see their Figure 13) have found that the star formation rate, beginning
$\sim$50 Myr ago, has been a factor of $\sim 20$ larger than the time-averaged rate 
for the galaxy, and therefore we illustrate the extent of the most vigorous star 
formation over the last 1 Gyr.  In Figure 7 we represent on the $V$-band image of the
galaxy four young populations of stars: 1) those stars with 
$-1.2 \lesssim U-B \lesssim -1.0$ and $V \lesssim 22$, which are the youngest, bluest
main sequence stars ({\it pluses}); 2) those young stars with magnitudes and colors, 
particularly in $U-B$, which make them likely main sequence turn-off stars and 
supergiants with ages $\lesssim 12$ Myr, including the very bright blue and red
supergiant stars mentioned in \S 3 ({\it crosses}); 3) those blue He-burning stars 
(with $-0.7 \lesssim U-B \lesssim 0.3$) with $V \lesssim 20.7$, i.e., those with
ages $\lesssim 50$ Myr ({\it circles}); and, 4) the corresponding red He-burning 
stars, seen on the $(I,V-I)$ CMD, with $1.2 \lesssim V-I \lesssim 1.6$ and 
$I \lesssim 19.7$ ({\it squares}).

As Dohm-Palmer et al.~(1997b) emphasize, the blue He-burning stars follow a
somewhat redder, nearly vertical track parallel to the main sequence and are about two
magnitudes brighter than the corresponding main sequence turnoff stars of the same
age.  The position in magnitude of a blue He-burning star on the CMD is determined
by the star's mass and, therefore, its age.  The blue He-burning stars are clearly
very useful age probes for stellar populations, back to about 600 Myr in
Sextans A.  This is only true at these low metallicities for blue He-burning stars
with ages older than $\sim$20 Myr, or $V \gtrsim 19.3$ for Sextans A, since the
He-burning phase for stars with $M > 15 M_{\sun}$ is not well-understood 
(\cite{chi92}; \cite{doh97b}).

One immediately can see that the recent star formation over the last 50 Myr has 
occurred in specific regions in Sextans A and is not uniformly distributed.  
In particular, the very young, hot, blue main sequence stars are primarily found near
surface brightness $\mu_B \simeq 25.9$ mag arcsec$^{-2}$, or a radius 
$r \simeq 2{\farcm}3$ (\cite{abl71}).  As pointed out previously by HSD, A87, and 
others, these hot stars are primarily found in the bright clusters in the southeast
and to the west.  But they are not exclusively in just these two locations.  Another 
significant clustering of these stars is to the northwest, and a small cluster is to the 
northeast.  A smaller number of these stars are found within the main optical body of the 
galaxy.  Large areas of the galaxy clearly exist where no recent star formation has taken
place.

What is most intriguing is that some of these stars are also found {\it beyond\/} 
$\mu_B \simeq 25.9$ mag arcsec$^{-2}$ ($r \simeq 2{\farcm}3$),
in the northeast and the southwest.  Based on Table 2, we consider the 
likelihood of foreground stars in such numbers superposed on this small field to be very
small.  We conclude that recent star formation appears to even be occurring beyond the 
main optical surface brightness of Sextans A.  Massive stars may form seemingly
``isolated;'' though this does not occur often, it is not unusual (e.g., Massey, Johnson,
\& DeGioia-Eastwood 1995; Massey et al.~1995).

The bright, young supergiants, not surprisingly, have a very similar spatial distribution 
to the very young main sequence stars.  (One supergiant star can be seen in the extreme
southeast, well beyond $r \simeq 2{\farcm}3$; it is possible that this is a 
foreground star, based on Table 2, although its magnitude and color [$V=18.92$, 
$B-V=0.52$, $U-B=0.31$] are consistent with other bright supergiants seen within this 
radius.)  But these stars, as plotted on Figure 7, have 
ages up to $\sim$12 Myr, and some are therefore older than the youngest population 
represented (the hot main sequence).  Some of these supergiants are seen in regions of the
galaxy not directly associated with the youngest stars, and therefore likely trace out 
regions of somewhat older star formation.

The evolved He-burning stars, both blue and red, with ages up to $\sim$50 Myr are also
found near $r \simeq 2{\farcm}3$, but, particularly in the west, their spatial 
distribution is appreciably inward, located more toward the galaxy's optical center, 
extending from the northwest down to the southwest.  Many of these stars are also found 
in the bright cluster of stars in the southeast, indicating that star formation in that
region has been occurring for at least $\sim$50 Myr.  The brighter stars with ages $\sim$50 
Myr in this cluster appear to form a ``ring,'' with radius of $\sim$300 pc.  Several of 
these blue-loop stars are also found in the smaller cluster of stars in the northwest.  
But, most noticeably, only one bright blue He-burning star is found in the galaxy's 
second-largest star-forming region, in the 
west, and this star, from its position on the CMDs (with $V=19.23$, $B-V=0.12$, and 
$U-B=-0.03$), is possibly as young as $\sim 15$ Myr.  This region of star formation, 
which Dohm-Palmer et al.~(1997a, b) could not study, is clearly one of the youngest in 
the galaxy, along with the small clustering of hot stars in the northeast.  Star formation
in these regions could be as recent as $\lesssim 20$ Myr.

In summary, the recent massive star formation over the last 50 Myr, which has been
so pronounced in Sextans A, has been confined toward the edge of the main body of the 
galaxy, near $\mu_B \simeq 25.9$ mag arcsec$^{-2}$, and may have progressed even beyond 
this point.  In particular, in the west, it has been progressing outward from more central
regions.  Much of the recent star formation has been concentrated in
the large clustering of stars in the southeast, and to the northwest.  The youngest
large concentration of star formation has been in the clustering in the west.  Noticeable
large holes can be seen in the distribution of young stars.  No appreciable recent star 
formation has occurred at the optical center of the galaxy.

Next, we examine the spatial distribution of older populations in Sextans A.
In Figure 8 we again show the $V$ image of the galaxy.  Shown on this figure
are two populations of stars: 1) those blue He-burning stars with ages
$\sim 50$--100 Myr, with $V \gtrsim 20.7$ to $V \sim 22$ (the faintest blue He-burning
stars that we can confidently resolve; {\it squares}); and, 2) the corresponding red 
He-burning stars with $19.7 \gtrsim I \gtrsim 21.3$, with ages to $\sim 100$ Myr
({\it circles}).

What is most noticeable about the distribution of the stars with ages 50--100 Myr is the
more central concentration than is the case for the younger populations.  Some structure 
and clustering also appears in the distribution of these older stars, which does not
appears to be random.  We have used a large age bin ($\sim$50 Myr wide) for these stars,
so small timescale details in their distribution as a function of age are lost.  
Nonetheless, the distribution of this population illustrates the advanced stage of 
dissolution of parent OB associations.  Regions where these stars are missing are 
noticeable in the figure.  Specifically, relatively fewer of the He-burning stars with 
ages $>50$ Myr are found at the sites of the most recent star formation (the bright blue
clusters) to the east and west.  In fact, no such stars are seen in the cluster to the 
west, again, indicating that this region is very young.  Additionally, the older He-burning
stars near the large cluster of young stars in the southeast tend to be to the southwest 
edge of the cluster, indicating a possible age gradient in this large star complex.  

The He-burning stars appear to mostly extend from far north down to far south in the
galaxy, with apparent clustering toward the galaxy's optical center.  Some of these stars
appear along a putative bar (see, e.g., \cite{abl71}; A87), although they are clearly not
exclusive to that region, if it exists.  (Ables 1971 describes this bar-like feature
as being seen in the isophotes with position angle 141{\fdg}5; Skillman et al.~1988
describe a possible H~I bar with position angle 105\arcdeg.)  The integrated color of the
bar feature, which appears particularly in the isophotal photometry, is $B-V=0.22$, 
excluding the star complex in the southeast (\cite{abl71}), and, thus, should be a mix of 
both blue and red stars.  These evolved He-burning stars should therefore comprise the
brighter populations along the bar, although we do not consider the optical bar structure 
to be entirely obvious from the distribution of these populations in Figure 8.
A few of the red He-burning stars are also visible beyond the main optical galaxy; 
although their colors and magnitudes are quite consistent with those of other similar 
stars within the galaxy, these could well be foreground stars.  In summary, the 
distribution of these older stars is clearly different from that for the younger stars.

Finally, in Figure 9 we once again show the $V$ image with two older populations: 
1) RGB stars, at roughly $I \sim 22$ and $V-I \sim 1.1$, and some AGB stars in the red 
tangle, with ages between $\sim$100 and $\sim$600 Myr ({\it crosses}); and, 
2) those RGB stars and AGB stars with ages $\gtrsim$600 Myr (the older red tangle and 
red tail stars with ages possibly up to $\sim$3 Gyr; {\it pluses}).  We chose the age 
ranges of 100--600 Myr and $> 600$ Myr, since Dohm-Palmer et al.~(1997b) have found that 
the star formation rate between about 100--600 Myr ago was a factor of $\sim$6 higher 
than the time-averaged rate, likely declining for ages older than this range; this is 
illustrated in their Figure 14.  

No particular pattern or structure is obvious in the spatial distribution for the RGB
stars.  However, few of these stars are seen in or near the youngest star formation regions.
Again, we are using a large age bin to increase our statistics, and therefore we cannot 
describe small timescale variations in the distribution.  These stars have likely dispersed
from their birthplaces over an area of diameter $\sim$5\arcmin.  Notice the approximate 
similarity of the distribution of RGB stars that we find and that found by Dohm-Palmer et 
al.~(1997b; their Figure 16) for their limited field-of-view, especially the possible 
clustering of RGB stars just east of the center.

One interesting aspect of the distributions for these stars is that they appear to also be
found outside $\mu_B \simeq 25.9$ mag arcsec$^{-2}$.  Again, given Table 2, one might suspect 
that $\sim$15--30\% of these stars are merely in the foreground.  But, again, the colors and 
magnitudes for the stars found outside the main body of the galaxy are consistent with those
red giants well within the galaxy, and a large fraction therefore could be members of Sextans 
A.  These stars may comprise a more extended halo-like structure, which is also found for a
number of dwarf irregular galaxies (e.g., Minniti \& Zijlstra 1996).

The older RGB and AGB stars with ages $\gtrsim$600 Myr are also uniformly
distributed across and beyond the main body of the optical galaxy, with no apparent clustering
or structure.  Again, the stars seen beyond $\mu_B \simeq 25.9$ mag arcsec$^{-2}$ have colors
and magnitudes consistent with those within the galaxy, but probably up to about one-third of 
these stars could be in the foreground.  The quite homogeneous distribution is what one would 
expect for an older halo field population.  Hunter \& Plummer (1996) point out that star
formation in Sextans A must once have occurred further out in the galaxy than is seen today,
given the size of the Holmberg radius.  They also point out that in the southwest
they can trace faint stars as far out as 1.5 kpc from the center.

What we find for stars older than 100 Myr is consistent with the picture developed 
by Dohm-Palmer et al.~(1997b, 1998), in which star formation, at a significantly lower 
rate than the current rate, percolates in coherent regions and propagates to other 
neighboring regions within their field-of-study during the time range of $\sim$100--600 
Myr.  The rate is likely lower for ages older than 600 Myr, but the pattern and 
propagation of star formation appear from our data to have been similar in nature.  
In particular, Dohm-Palmer et al.~(1998) show that from 300--400 Myr, star
formation migrated from the northeast to the center, peaking there around 350 Myr ago.
Our results seem to imply that this centrally concentrated star formation continued to
peak from $\lesssim$100 Myr.

\section{Relationship of the Stars to the Gas}

We now examine the spatial distribution of the various stellar populations to
that of the gas in Sextans A.  We can consider only the ionized and neutral
components of the gas in the galaxy, since molecular gas has not been detected
(Ohta et al.~1993 found only a low upper limit for the luminosity of CO).

\subsection{Ionized Hydrogen}

First, we examine the distribution of blue stars relative to the ionized gas, as
seen from our H$\alpha$+[N II] image.  In Figure 10 we show our H$\alpha$ map, which
is qualitatively similar to that in Aparicio \& Rodriguez-Ulloa (1992) and Hodge, 
Kennicutt, \& Strobel (1994a).  All the H$\alpha$ emission appears to be non-stellar.
The H~II regions are distributed in two arcs separated
by nearly 4\arcmin\ (with the brightest patches of emission in the southeast).  Fainter
emission can also be seen.  Our H$\alpha$ image, 
although continuum-subtracted, is not flux-calibrated, but comparing our
map with that in Hodge et al.~(1994a), we estimate that our detection limit
for H~II regions is $\sim 2.5 \times 10^{-14}$ erg cm$^{-2}$
s$^{-1}$, which, at a distance of 1.4 Mpc, corresponds to a luminosity of
$\sim 6 \times 10^{36}$ erg s$^{-1}$.  Some of the fainter structures seen on
the deep H$\alpha$ map in Hunter \& Gallagher (1997), such as the long filaments 
(e.g., their filaments 1 and 2) and fainter shells (e.g., their shell 4) are just barely
visible on our map and are somewhat below our detection limit.

In Figure 10 we show the blue main sequence stars (as in Figure 7,
with $-1.2 \lesssim U-B \lesssim -1.0$ and $V \lesssim 22$) relative to the
H$\alpha$ emission.  These stars are roughly spectral type B0 and hotter, and 
$\lesssim 8$--10 Myr old (at the assumed metallicity of Sextans A, this 
corresponds to a main sequence turnoff mass of $\sim$20 $M_{\sun}$).  Stars
cooler than this do not contribute significantly to the UV continuum flux 
which can produce H~II regions in the galaxy.  One can see that the hot, 
massive stars, as expected, are well-correlated with the ionized gas, both bright
and faint (in fact, a cluster of six stars cannot be distinguished from the dark
contrast in Figure 10 for the H~II region ``no.~17,'' following the Hodge et 
al.~numbering scheme, or ``no.~7,'' following that of \cite{apa92}).  Thus,
we have accounted for the ionizing sources for the majority of the H~II gas.

A lesser number of presumed OB stars are not associated with any H~II regions
above the detection limit of our study.  This is true also for the regions detected
by Aparicio \& Rodriguez-Ulloa (1992) and Hodge et al.~(1994a), although
some appear to be associated with some of the fainter emission seen in the
Hunter \& Gallagher (1997) map.  Additionally, some of the fainter emission
features, based on closer inspection of our images, may be ionized by unresolved
small clusters of hot stars, as is also likely the case for the bright knotty 
emission region, which is Hodge et al.~(1994a) H~II region ``no.~20'' and Aparicio 
\& Rodriguez-Ulloa (1992) ``no.~2''.
However, the inner area of the galaxy is relatively devoid of massive
stars and H~II regions (the inner portion of the galaxy may have several
faint filaments of emission, and, as Hunter \& Gallagher 1997 point out, 
these filaments might comprise a larger shell structure associated with the
bright H~II complex in the southeast).  The bulk of the massive star formation
within the last $\sim 20$ Myr has been spatially, and temporally, correlated 
along the edge of the main surface brightness for the galaxy.

\subsection{Neutral Hydrogen}

Next, we investigate the relationship of the stars to the neutral gas.
We have utilized a radio 21 cm H~I map from Graham \& Westpfahl (1998).  
Here we dispense with the details on their H~I data, only 
briefly to indicate that this H~I map was made using the Very Large Array
(VLA) of the National Radio Astronomy Observatory\footnote{The National Radio
Astronomy Observatory is operated by Associated Universities, Inc., under
cooperative agreement with the National Science Foundation.} at the redshifted
frequency of the 21 cm line of neutral hydrogen appropriate for the galaxy,
in the C-configuration in 1992 May and the D-configuration in 1992 July.
(B-configuration observations were also made in 1994 June, but are not included
in the map shown here.)
These two sets of data were then combined, resulting in a naturally-weighted
map with resolution $34\arcsec \times 23\arcsec$ and rms noise in the channel
maps of 1.36 mJy beam$^{-1}$.  Both the resolution and the sensitivity are
higher than those for the H~I map presented in Skillman et al.~(1988).  Additionally,
with the short spacings sampled in the most compact configuration, the maps
are sensitive to more extended emission than were the Skillman et al.~data.
(An initial analysis of the data by Puche, including moment maps, can be found 
at the URL http://cfa-www.harvard.edu/projects/nga/sexa.html.)

In Figure 11 we show the total H~I column density map for Sextans A.  
What Puche \& Westpfahl (1994) consider a central depression, 
or ``hole,'' within an H~I shell can clearly be seen; Skillman et al.~(1988)
describe the H~I morphology as a ``horseshoe or ring-like'' structure around a
central minimum, with a sharp outer boundary.  This hole, $\sim$1.5 
kpc in diameter, encompasses nearly the entire main optical surface brightness
of the galaxy.  The highest
column density forms the two large spur- or arc-like structures toward the southeast 
(peak column, $\sim 1.6 \times 10^{22}$ cm$^{-2}$) and the northwest
(peak column, $\sim 1.2 \times 10^{22}$ cm$^{-2}$), as also seen in the map by
Skillman et al.~(1988), and highlight the nearly complete ring or shell.  The column
density in the center of the galaxy, in the hole, is $\sim 1.2 \times 10^{21}$ cm$^{-2}$.
The fine detail in this map indicates substructuring of smaller, fainter clumps,
or clouds, and holes within the larger H~I hole as well.  
Additionally, one can see that H~I extends well beyond the H~I ring or shell, as well
as the main optical galaxy, consistent with other normal galaxies and what van Zee, 
Haynes, \& Giovanelli (1996) found for their sample of dwarf galaxies.  A fascinating 
feature of the map is the ``hook'' of gas extending from the disk to the west.  Again, 
we defer here to Graham \& Westpfahl (1998) for a complete presentation of these H~I data.

In Figure 11 we also show the positions of the same blue stars as in Figure 10 
({\it pluses}).  What should be kept in mind is that the gas disk is likely thicker than
that for spirals (\cite{puc92}) and that the H~I shell is roughly spherical.
What is most striking is that the bluest, youngest stars are found preferentially and 
predominantly along the {\it inner\/} edge of the large H~I hole, particularly along the 
inner edge of, and not coincident with, the two large complexes of H~I gas to
the southeast and the northwest.  Far fewer of these stars are seen within or outside 
of this locus.  

In Figure 12 one can also see, not unexpectedly, that the H$\alpha$ emission is along the
inner edges of the H~I hole.  The two large H~II region complexes, in the southeast and the
northwest, are not coincident with the brightest H~I peaks, but, in fact, straddle the 
{\it inner\/} edge of both peaks.  Hunter \& Plummer (1996) noted that the distribution 
of H$\alpha$ does not extend as far as the H~I gas distribution.  This is consistent with
the fact that the most recent star formation is occurring primarily along the inner edge of
the H~I hole.  

In Figure 13 we show the distributions of the young, blue main sequence stars, supergiants
and He-burning stars together, as in Figure 7.  One can see that the overall spatial 
distribution of stars with ages up to $\sim$50 Myr appears to have a similar shape and 
symmetry axis as the current distribution of the H~I gas, and the youngest stars, but with
a smaller radius.  If the peak velocity dispersion in the H~I 
($\sim$9--10 km s$^{-1}$; \cite{ski88}; \cite{gra98}) represents a possible expansion of 
the gas outward to form the large H~I hole, then to produce the current size of the hole 
($\sim$1.5 kpc) would take $\sim$80 Myr, assuming a constant expansion rate.  If stars have
been consistently forming along the inner edge of the H~I hole as the gas homologously 
expanded, then we might expect stars with ages up to $\sim$50 Myr to share a similarly
ring-shaped spatial distribution as the most recently-formed stars, albeit closer to the 
galaxy's center, since $\sim$50 Myr ago the H~I hole would have been $\sim$0.5 kpc smaller 
in radius.  (However,  OB associations would have dispersed somewhat over a $\sim$50 Myr 
timescale)

In Figure 14 we show the distributions of the older He-burning stars, as
in Figure 8, with ages $\sim$50--100 Myr.  What is immediately evident is that
these older stars are concentrated more toward the center of the H~I hole and bear 
relatively little relationship with the H~I hole, shell, and substructuring than do 
the younger stars in Figure 13.  We point out that regions within the H~I hole exist 
where no stars from these older generations appear to have formed.  It is clear
that the gas must have been more centrally concentrated in the past than it is now.

Finally, in Figure 15 we show the distributions of the older RGB and AGB stars,
as in Figure 9, with ages $\sim$100 Myr -- 1 Gyr or so.  Here we see no relationship of 
these stars to the current structure of the neutral gas, consistent with the substantial
dispersal of these stars from their birthplaces.  The populations of older stars appear 
nearly uniformly distributed relative to the observed H~I.  
Although our optically imaged field encompasses nearly the entire H~I hole, it does not 
overlap much with the gas beyond the H~I shell.  The indications in Figure 15, however, are 
that stars had formed long ago well beyond the current H~I shell structure.  The H~I gas 
in the extended disk outside the shell implies that the star-forming molecular gas may 
well have existed (and may still exist) there.

\section{Discussion}

What could have been responsible for the substantial increase in the rate and spatial 
concentration of the recent star formation in Sextans A?  The answer to this puzzle has 
ramifications both for this particular galaxy and for the bursts of star formation occurring
in other dwarf irregular galaxies.  Dohm-Palmer et al.~(1997b) conclude that sometime within
the last 100 Myr or less, an event, or events, transpired that launched the galaxy into its 
present star-forming high state (although the movie in \cite{doh98} might imply that this 
event occurred as early as $\sim$370 Myr ago).  We briefly speculate 
here on some of the possible mechanisms.  A tidal interaction with nearby Sextans B
has been dismissed above as a possibility, due to the large separations between the two
galaxies.

One possibility is the presence of a bar potential in the galaxy.  Several
authors have stressed the existence of a possible bar-like feature, evident both in the
optical isophotal data (\cite{abl71}) and from the gas kinematics ({\cite{ski88}).  A misalignment
of the H~I morphological axis with the kinematical axis, and, possibly, with the optical 
morphology, is typically an indication of a bar.  The bar feature appears somewhat off-center, 
as is true for Magellanic irregulars more massive than Sextans A.  The H~I bar presumably 
connects the two pronounced maxima in the H~I distribution and may be associated with the gas 
pile-up at its ends, where the most prominent recent star formation is occurring.  

Although less is known about bars in dwarf irregular galaxies than for other larger galaxies, 
we can make inferences based on what is known for the more massive systems.  Bars in 
irregulars (of generally greater mass than Sextans A) are common and may possibly arise from 
interaction with small neighboring H~I clouds (\cite{wil96}).  A bar can form in a disk in about
one-half to one rotation period as a result of a merger or tidal interaction (\cite{nog87}; 
\cite{bar91}).  Additionally, late-type galaxies tend to have H~I gas-rich bars (\cite{hun96c}),
with star formation occurring along the bar (\cite{phi96}), which appears to be the trend for,
e.g., the Magellanic irregulars NGC 4618 (\cite{ode91}) and NGC 4449 (\cite{hun96b}).

The main difficulties with a bar explanation for Sextans A are the distribution of gas and
star formation in the galaxy and the necessary timescale for bar evolution.
Induced inflow of gas along a bar (e.g., \cite{sch84}) could account for the fact that the
gas in Sextans A was once more centrally concentrated $\gtrsim$50 Myr ago.  However, currently 
the H~I gas, and star formation, is concentrated in a shell around a central hole.
The evolution of a bar, in the absence of interaction, likely requires several rotation periods.
However, a bar can be destroyed in about one rotation period, with the formation of a long-lived 
outer gaseous ring, after a tidal interaction or a central accumulation of mass (\cite{ath96}).
Redistribution of the gas from the bar into a ring could explain the current gas distribution.
Yet, the dynamical timescale for Sextans A is longer than these model bar evolutionary 
timescales, since the galaxy has a rotation period $\sim 400 (\sin i)$ Myr (\cite{ski88}).  

Furthermore, we have isolated the region of the galaxy thought to be the putative optical bar 
(cf.~\cite{abl71}) and find that the ages of the bar stars are typically $\gtrsim$40--50 Myr old.
We find no difference, then, between the ages of these stars and those of the stars generally 
throughout the central regions of the galaxy.  The presence of an optical bar is therefore not 
convincing; we may merely be seeing the remnants of an extended OB association or 
associations, analogous to Shapley Constellation III and environs in the LMC.

A second possibility is the effect of a merger or accretion event in the past $\sim$100 Myr (with
or without bar formation).  This is the mechanism supported for Sextans A by Graham \& Westpfahl 
(1998).  Evidence for merger events appears to exist for a number of dwarf galaxies.  Saito et 
al.~(1992) propose such a model for IC 10.  Of particular curiosity in the case of Sextans A is 
the H~I gas extending westward from the H~I disk.  Graham \& Westpfahl claim that the velocity
field for the galaxy can be explained as resulting from a merger.  Mergers are also known to 
result in a central concentration of the molecular gas in galaxies.
Thus, a merger model cannot be easily dismissed.  However, it alone does not offer an explanation
for the current distribution of star formation in Sextans A. 

A third possibility is the mechanism of supernova-driven expansion and sequential star 
formation, at least for more recent epochs.  Hunter et al.~(1998) have recently 
suggested that sequentially-triggered star formation, driven by mechanical energy input by 
concentrations of massive stars, is likely to be an important mechanism for cloud formation in 
irregulars.  The fact that the current star formation in Sextans A is occurring mostly on the 
edge of the H~I hole is consistent with cold gas pileup and compression at this locus.
This would occur if the gas within the H~I hole is expanding outward into the extended H~I 
envelope seen in the radio data.  The behavior of the current star formation is remarkably 
similar to the star formation occurring particularly along the largest H~I holes seen in Ho~II 
(\cite{puc92}; see also \cite{ton95}) and IC 2574 (Martimbeau et al.~1994).  A local example
may also be Constellation III and LMC 4 in the LMC (\cite{dop85}; \cite{gre98}).  

As massive stellar populations have formed and evolved in the past in the interior of Sextans A, 
they have input significant energy into the ISM through stellar winds (e.g., Leitherer et 
al.~1992) and ended their lives as supernovae.  The resulting hot gas expands outward from the 
interior of the galaxy, pushing on the colder gas, compressing it, forming clouds, and triggering
further star formation as the galaxy evolves.

Is this mechanism plausible for Sextans A, given the kinematics of the gas?  
As discussed in \S 5.2, if the peak velocity dispersion in the H~I ($\sim$9--10 km s$^{-1}$;
\cite{ski88}; \cite{gra98}) represents a possible expansion of the gas to form the large H~I 
hole, then the age of the hole would be $\sim$80 Myr.
This timescale is consistent with the fact that the star 
formation rate substantially increased in the last 100 Myr and also agrees with the ages of stars
near the center of Sextans A, which would presumably be the remaining less massive siblings of 
the supernova progenitors from those stellar generations.  Furthermore, the distributions of 
populations with ages $<50$ Myr is suggestive of a progression from oldest stars inward to the 
youngest outward within the H~I hole, consistent with an expanding front of star formation
outward from the center.  

A major difficulty with this scenario is that, although Puche \& Westpfahl (1994) claim evidence
for expanding gas, Graham \& Westpfahl (1998), from the very same dataset, do not find convincing
evidence for an expanding shell, i.e., characteristic double-peaked velocity profiles, 
particularly at the velocity dispersion peak, which they find near the galaxy's center, in the 
lowest column density region.  However, it may be difficult to see such a feature in these low 
brightness regions.  Furthermore, Domgoergen, Bomans, \& de Boer (1995) and Points et al.~(1998)
found an absence of expansion for LMC 4 and for LMC 2, respectively.

Another potential difficulty is that Rhode, Salzer, \& Westpfahl (1998), from deep optical CCD 
imaging, recently claim not to find remnant populations at the centers of H~I holes in Ho~II
and other dwarfs.  However, Rhode et al.~searched for remnant main sequence populations at the 
hole centers, whereas it is clear from our CMDs that the remnant populations would be most readily
identified as evolved He-burning stars, which are $\sim$2 mag brighter, and significantly redder, 
than main sequence stars of the same age.  (Interestingly, red supergiants might dominate the 
light from a remnant cluster, since, in systems with the low metallicities of dwarf irregulars, 
these stars may not necessarily be the progenitors of all massive supernovae, as we have directly 
witnessed with SN 1987A in the LMC; see \cite{woo88}, \cite{sai88}.)
For Sextans A, we believe that our imaging shows several populations of a feasible age, near the 
center of the large hole, which are suitable candidates for remnant sibling stars to the 
supernovae progenitors.

Of course, some combination of these various mechanisms may have been at work during the galaxy's
history.  It is possible that the galaxy has experienced the effects of a merger or accretion 
event, or of a bar potential sometime in the past, resulting in a significant increase in the 
star formation rate.  The increased massive star formation resulted in increased feedback to the 
ISM, which may have resulted in an expanding bubble of hot gas, which has promoted the current 
star formation we are witnessing.  We simply do not have enough information.  

The trigger for the current burst of star formation in Sextans A remains a mystery, although
we have shed additional new light through our optical imaging of the galaxy and our
comparison with the gas components.  Dohm-Palmer et al.~(1997a,b) have also provided some vital
clues, but due to the limited field size of the HST/WFPC2 camera, they sampled only about half of 
the galaxy, and therefore, their results are only applicable to this smaller field.  Since
star formation has occurred in bursts in various regions of Sextans A, sampling a larger area of
the galaxy provides a broader view of the recent star formation history.

Unfortunately, one important additional clue that is missing is the distribution of the molecular
gas, about which we have no information other than the low upper limit to the luminosity of the CO
tracer (\cite{oht93}).  We do not know the mass nor the extent of the actual gaseous raw material for 
star formation in the galaxy.  That information would help resolve questions about the nature and
history of star formation in Sextans A.

\section{Conclusions}

From our ground-based observations of the Local Group dwarf irregular galaxy Sextans A in 
$UBV(RI)_{\rm C}$ and H$\alpha$, we have found that current star formation is occurring mostly 
along the inner edge of the large H~I shell surrounding a hole, or depression, in the H~I column 
density, consistent with cold gas pileup and compression of the gas at this locus.  The distribution
of stars with ages $\lesssim$50 Myr indicate an outward progression of star formation with time.
Regions exist where the star formation appears to be exclusively recent, and regions where no star
formation has occurred within the last 50 Myr.  Older star formation, from $\sim$50--100 Myr, 
appears to have some spatially coherent structure and is more centrally concentrated than the most
recent star formation.  The oldest star formation that we can accurately sample appears to be 
uniformly distributed across Sextans A, even beyond a surface brightness of $\mu_B \simeq 25.9$ 
mag arcsec$^{-2}$, or, a radius $r \simeq 2{\farcm}3$.

The recent bursts in star formation are not only coherent in time, but also in space across Sextans
A.  The trigger of these bursts remains unknown, however, we believe we have presented
the strongest evidence so far that supernova-driven expansion of the gas may be responsible for at
least some of the star formation over the galaxy's history.  The expansion of the gas, due to episodic
sequential star formation within the galaxy, is a direct trigger for further star formation as the 
galaxy evolves.  We note that the expected expansion age of the H~I shell is $\sim$80 Myr 
(assuming an expansion velocity of $\sim$10 km s$^{-1}$ and a shell diameter of $\sim$1.5 kpc), 
which is consistent with the ages of the older, centrally concentrated evolved stars, which may
be the remnants of the populations which included the massive stars responsible for the expansion.

\acknowledgements

We are grateful to Ralph Young Shuping for his photometric calibration observations
of Sextans A at the KPNO 0.9-m telescope, to NOAO director Sydney Wolff
for scheduling these observations, and Dave Summers for his assistance at the
KPNO 2.1-m.  We thank Maggie Graham and 
Dave Westpfahl for making their total H~I map available to us.
SVD is grateful for assistance provided by
Alex Filippenko and also to the UCLA Division of Astronomy and Astrophysics.
We thank Jean Turner, Tammy Smecker-Hane, Dave Westpfahl, and Dave Meier for
helpful discussions.  We also thank the referee for very helpful comments and suggestions.
This research was supported by a grant from NASA administered by the American
Astronomical Society.

\clearpage

\begin{deluxetable}{ccc}
\def\phmm{\phm{$-$}}
\tablenum{1}
\tablecolumns{3}
\tablewidth{5in}
\tablecaption{Comparison with Photometry from Previous Studies}
\tablehead{
\colhead{Authors} & \colhead{Source} & \colhead{Difference}}
\startdata
\cutinhead{$V$}
HSD & CCD $gri$ & \phmm$0.350 \pm 1.024$ \nl
SC85 & photog $BV$ & \phmm$0.533 \pm 0.711$ \nl
A87 & CCD $UBV$ & $-0.025 \pm 0.494$ \nl
W87 & CCD $BV$ & \phmm$0.013 \pm 0.098$ \nl
SMF & CCD $BVI$ & \phmm$0.035 \pm 0.031$ \nl
\nl
\cutinhead{$B-V$}
HSD & CCD $gri$ & $-0.140 \pm 0.516$ \nl
SC85 & photog $BV$ & \phmm$0.185 \pm 0.251$ \nl
A87 & CCD $UBV$ & $-0.125 \pm 0.443$ \nl
W87 & CCD $BV$ & \phmm$0.042 \pm 0.082$ \nl
\nl
\cutinhead{$U-B$}
A87 & CCD $UBV$ & \phmm$0.076 \pm 0.438$ \nl
\nl
\cutinhead{$V-I$}
SMF & CCD $BVI$ & \phmm$0.046 \pm 0.054$ \nl
\nl
\cutinhead{$R-I$}
HSD & CCD $gri$ & \phmm$0.093 \pm 0.198$ \nl
\enddata
\end{deluxetable}

\clearpage

\begin{deluxetable}{cccc}
\def\phmm{\phm{$-$}}
\tablenum{2}
\tablecolumns{4}
\tablewidth{6in}
\tablecaption{Comparison of Actual and Predicted Foreground Star Counts}
\tablehead{
\colhead{$V$} & \colhead{$B-V<0.8$} & \colhead{$0.8<B-V<1.3$} & \colhead{$1.3<B-V$}}
\startdata
\cutinhead{Actual Counts}
18 & 8 & 1 & 4 \nl 
20 & 196 & 20 & 40 \nl
22 & 725 & 472 & 83 \nl
24 & 55 & 209 & 54 \nl
\nl
\cutinhead{Predicted Counts\tablenotemark{a}}
18 & 1.1 & 1.4 & 1.9 \nl
20 & 2.4 & 1.1 & 5.5 \nl
22 & 2.4 & 2.4 & 10.8 \nl 
24 & 0.9 & 2.8 & 17.7 \nl
\enddata
\tablenotetext{a}{Based on interpolating the model counts from Ratnatunga \& Bahcall 
(1985) for the field appropriate for Sextans A.}
\end{deluxetable}

\clearpage

\begin{figure}
\figurenum{1}
%\plotone{vandyk.fig1.ps}
\caption{A composite three-color ($UV[I]_C$) image 
of Sextans A, with north up and east to the left.}
\end{figure}

\clearpage

\begin{figure}
\figurenum{2}
\plotone{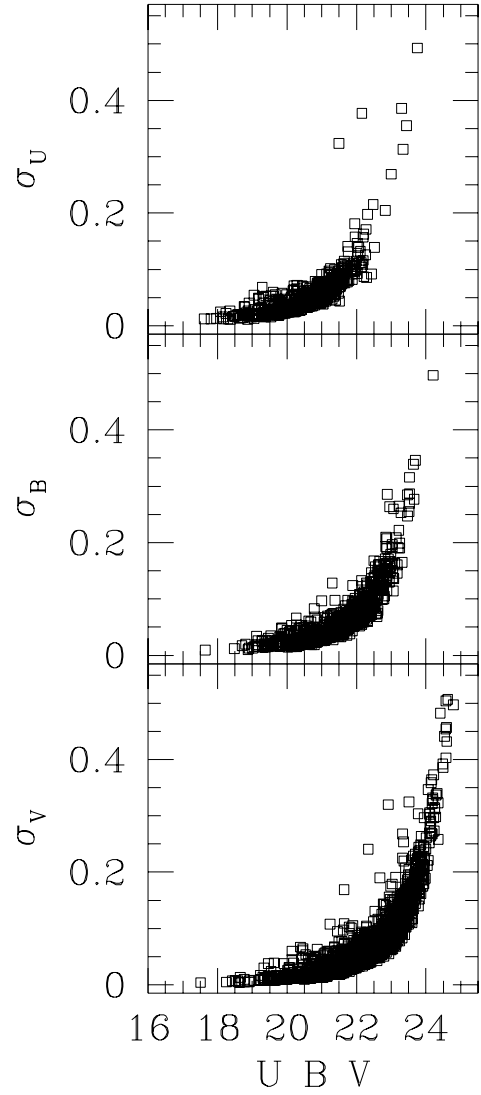}
\caption{Distribution as a function of magnitude of the standard errors derived
by ALLSTAR in DAOPHOT for the $UBV(RI)_C$ bands.}
\end{figure}

\clearpage

\begin{figure}
\figurenum{2}
\plotone{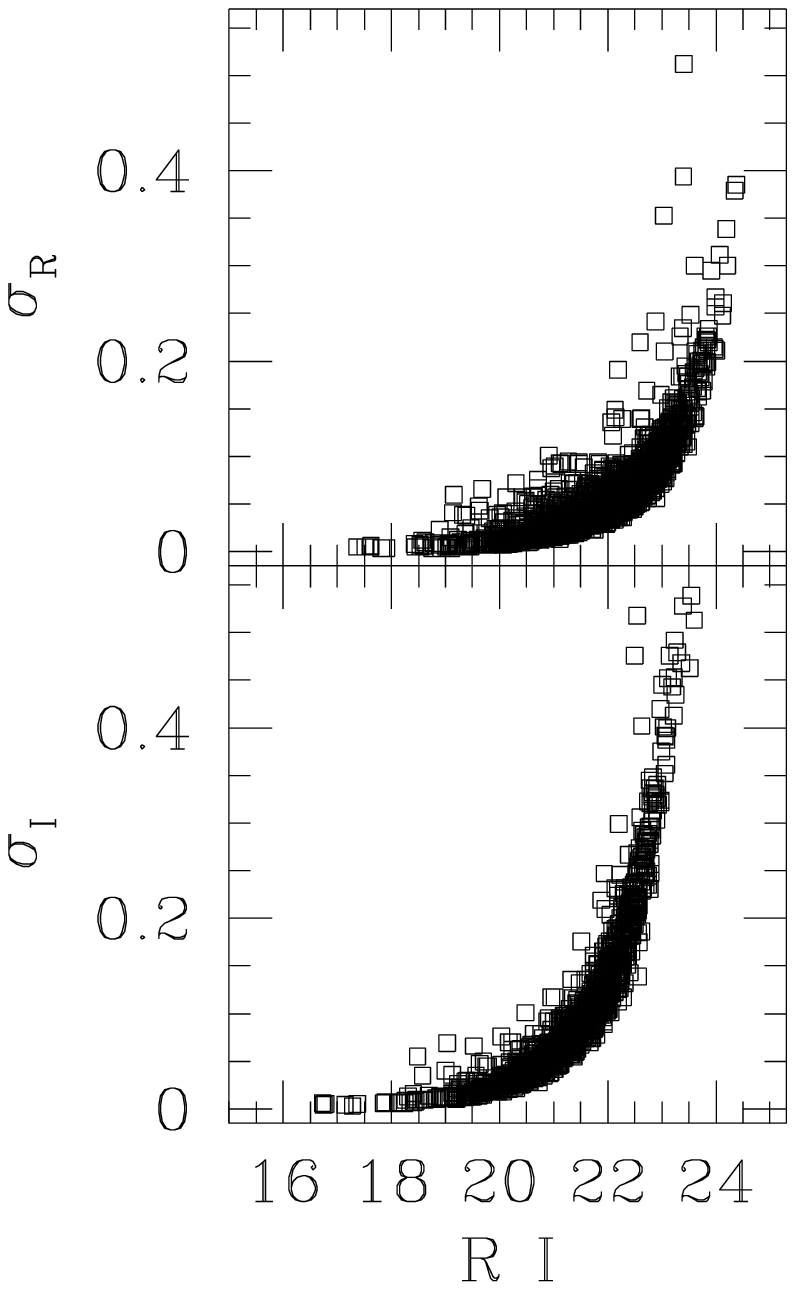}
\caption{(continued)}
\end{figure}

\clearpage

\begin{figure}
\figurenum{3}
\plotone{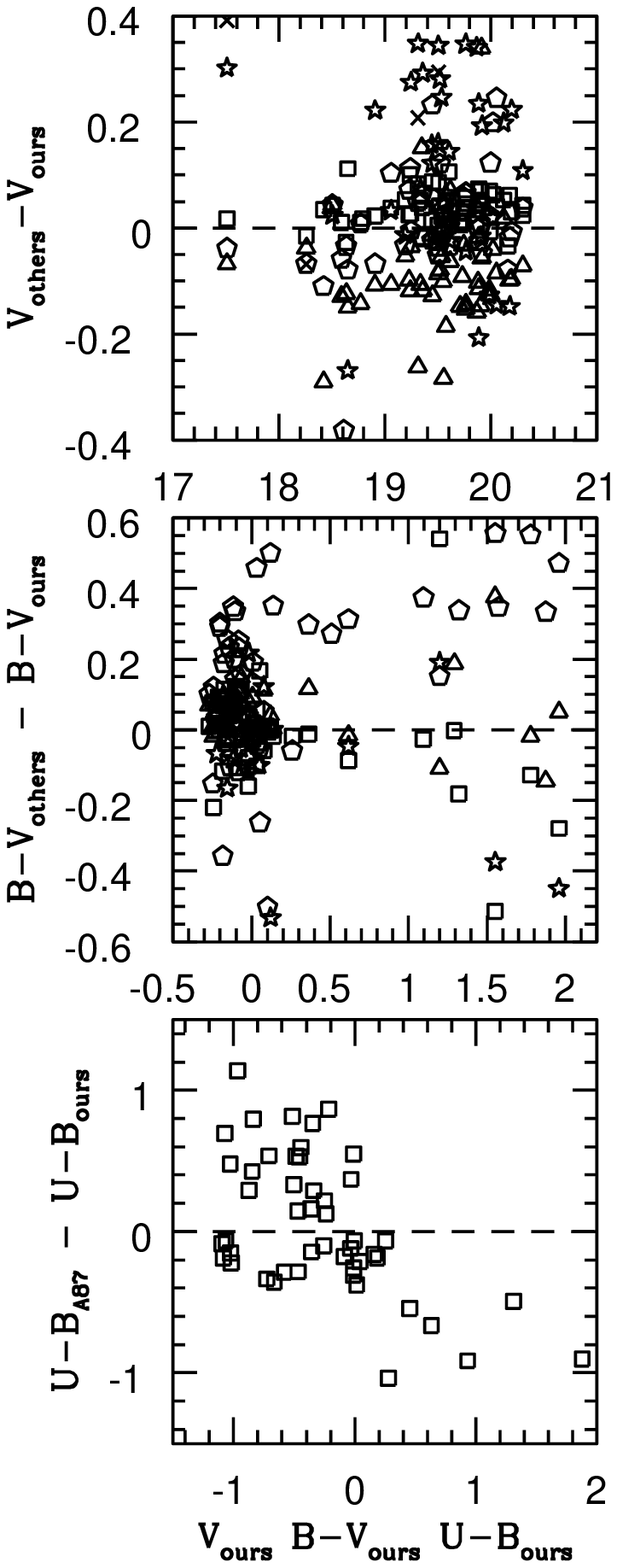}
\caption{Comparison of our photometry with the photometry from other investigators
in the various bands.  See text for author abbreviations.
The symbol code for $V$: {\it open squares}=SMF; 
{\it open triangles}=A87; {\it open pentagons}=W87; {\it stars}=SC85;
{\it crosses}=HSD. 
The symbol code for $B-V$: {\it open squares}=A87; 
{\it open triangles}=W87; {\it open pentagons}=SC85; {\it stars}=HSD.
The symbol code for $U-B$: {\it open squares}=A87.
The symbol code for $R-I$: {\it open squares}=HSD.
The symbol code for $V-I$: {\it open squares}=SMF.}
\end{figure}

\clearpage

\begin{figure}
\figurenum{3}
\plotone{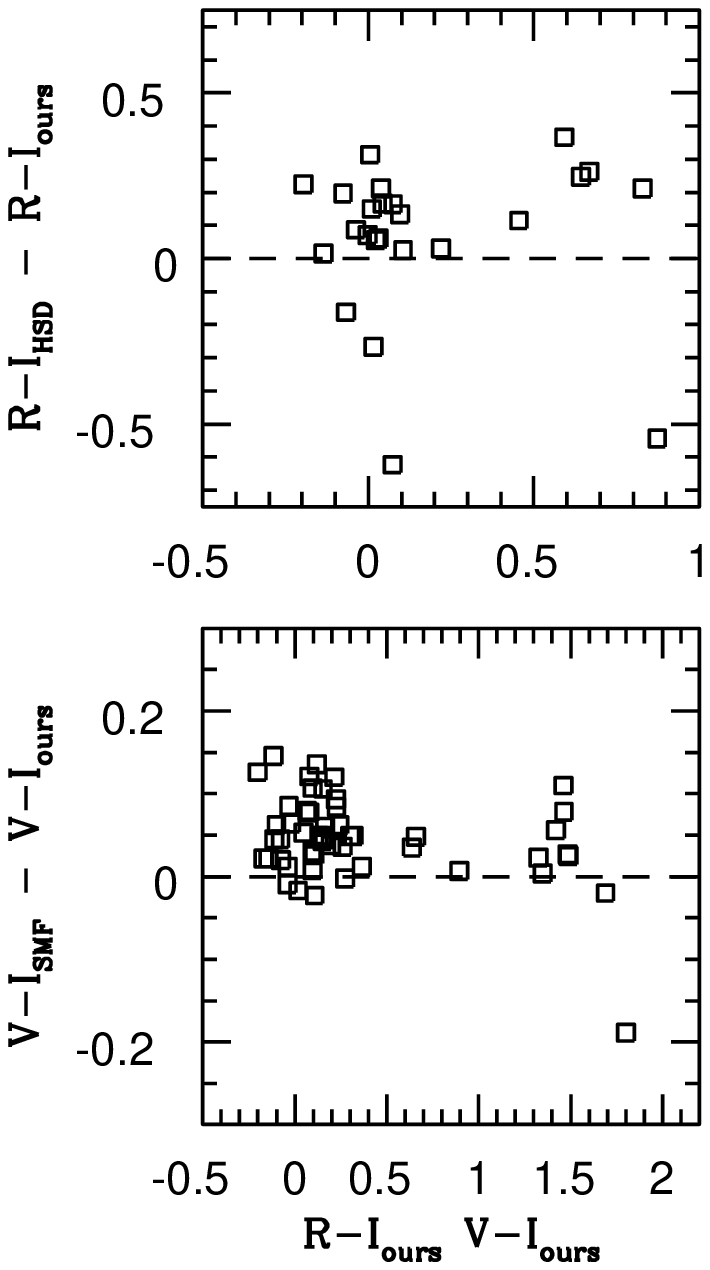}
\caption{(continued)}
\end{figure}

\clearpage

\begin{figure}
\figurenum{4}
\plotfiddle{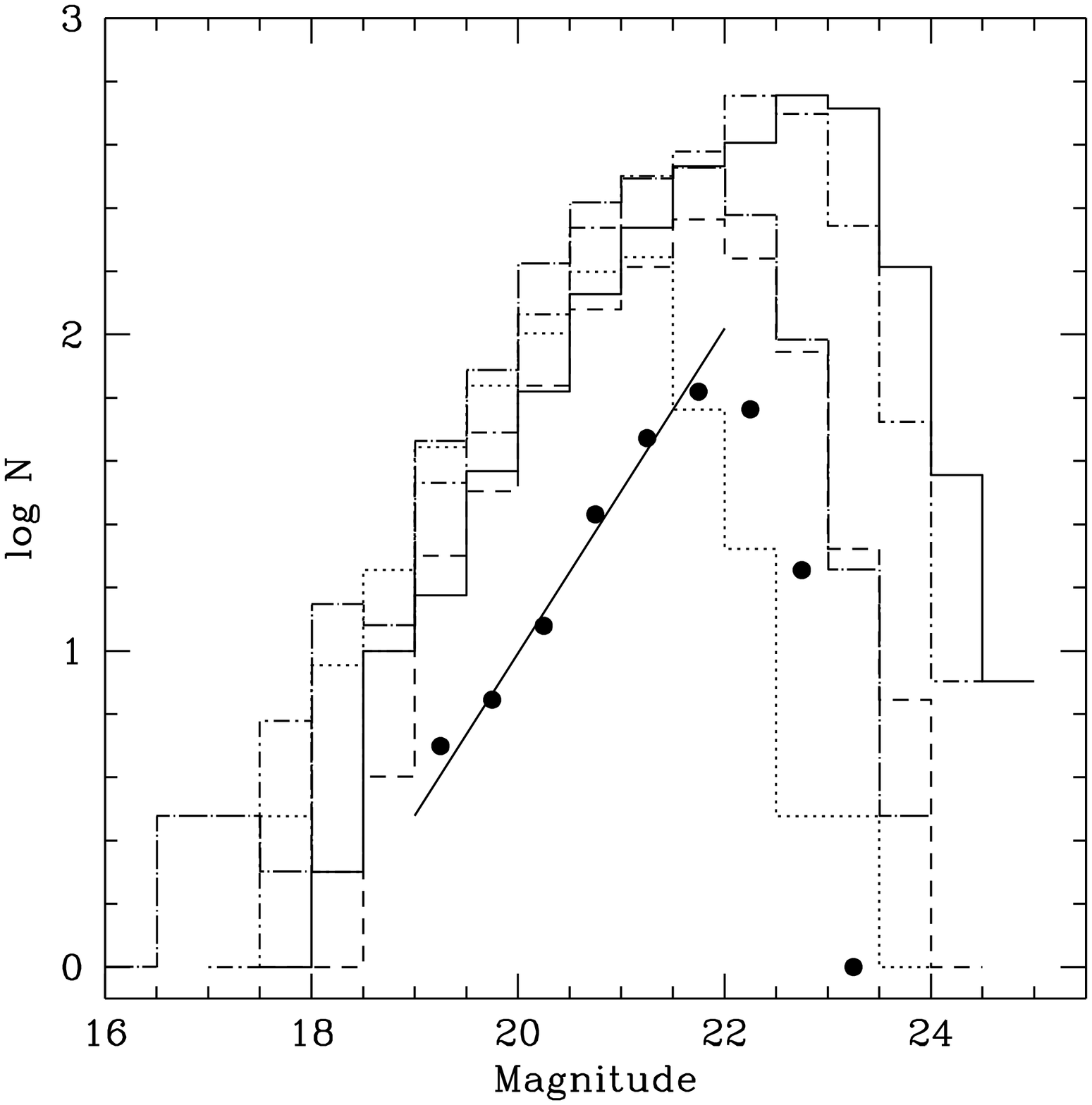}{300pt}{-90}{65}{65}{-250}{+400}
\caption{Differential luminosity functions for Sextans A in 
$U$ ({\it dotted line}), $B$ ({\it dashed line}),
$V$ ({\it solid line}), $R$ ({\it short dashed-dotted line})
and $I$ ({\it long dashed-dotted line}), for all detected stars.  
Also shown is the function for only main sequence stars {\it filled
circles}, i.e., those with $U-V \lesssim -1.05$.  The line represents a 
least-squares fit to the main sequence points for stars with 
$V \lesssim 22$, resulting in a slope $d\log N / dV = 0.48$, which is
consistent with that found for other dwarf irregular galaxies, and also
with the functions for all stars in all the bands.}
\end{figure}

\clearpage

\begin{figure}
\figurenum{5}
\plotone{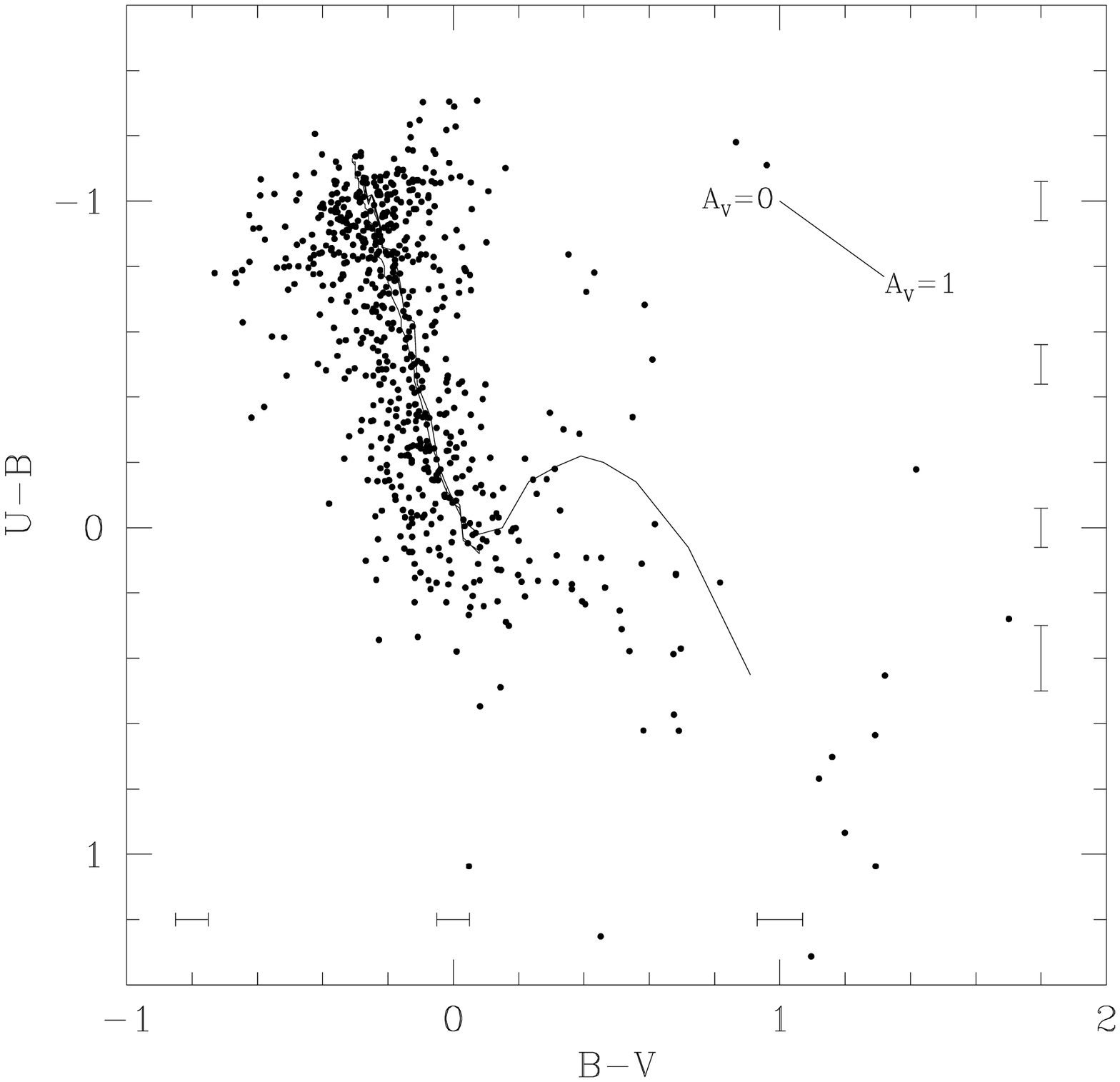}
\caption{The color-color diagrams for Sextans A.  In {\it a)} we show
the $(U-B, B-V)$ diagram, and in {\it b)} we show the 
$(B-V, V-I)$ diagram.  On each diagram we show the unreddened 4 Myr isochrone
from Bertelli et al.~(1994) with metallicity $Z=0.001$; 
on {\it b)} we also show the 100 Myr isochrone
(note the poorer agreement of the AGB stars with the model).  
On each we also show representative uncertainties in the colors and
the direction of the reddening vector.}
\end{figure}

\clearpage

\begin{figure}
\figurenum{5}
\plotone{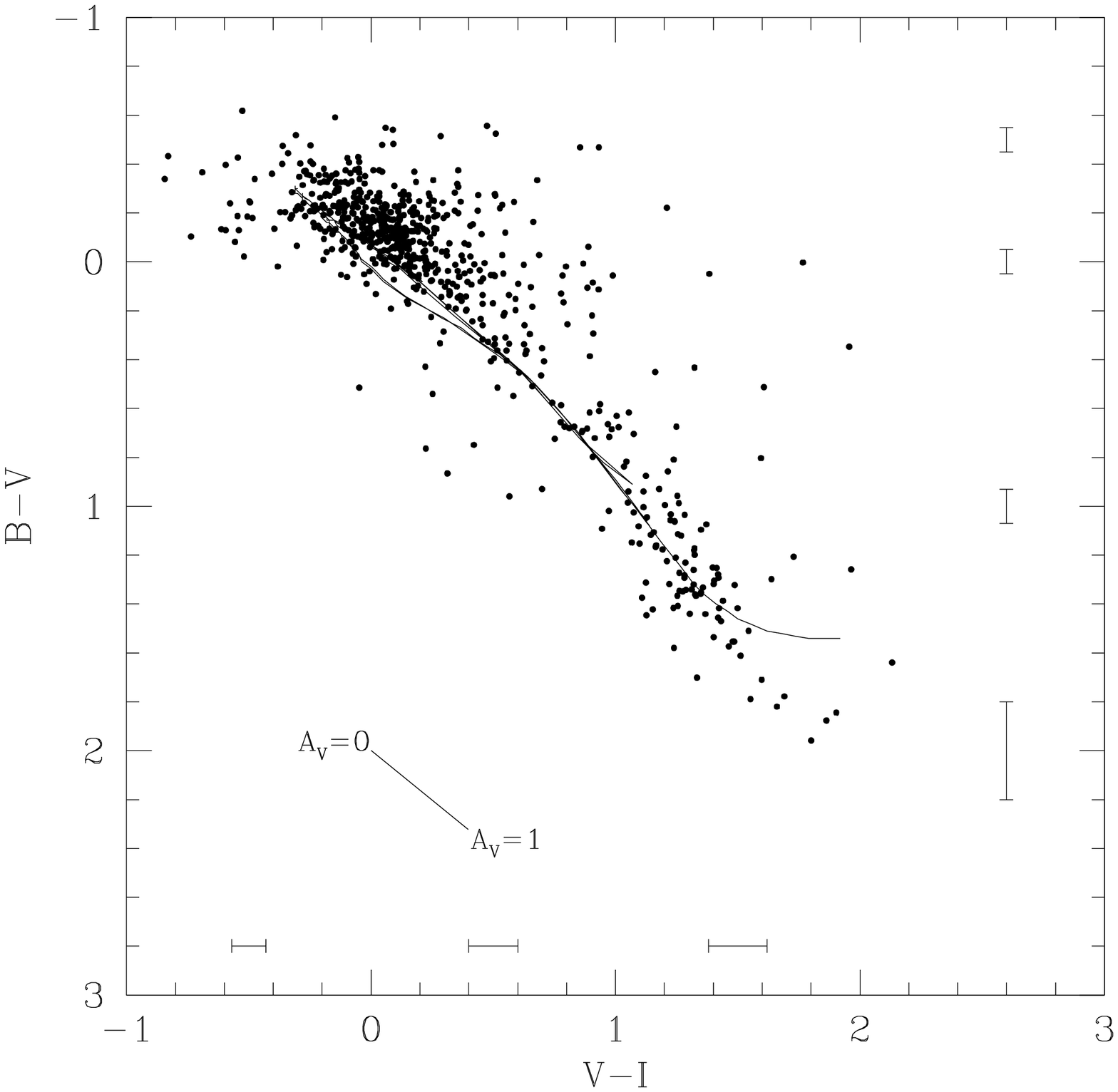}
\caption{(continued)}
\end{figure}

\clearpage

\begin{figure}
\figurenum{6}
\plotone{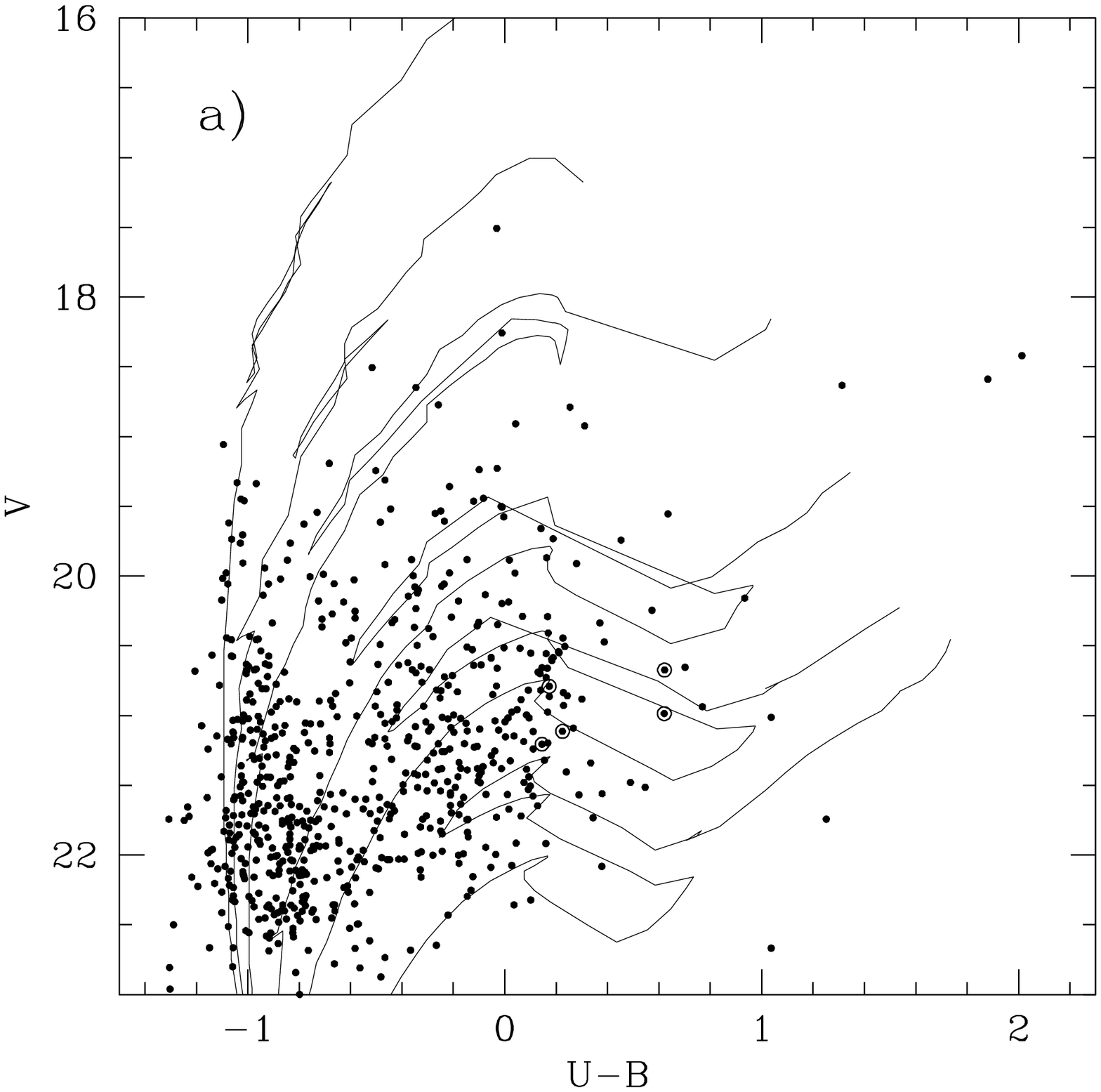}
\caption{Color-magnitude diagrams for the stars in Sextans A.  On {\it a)}
we have also plotted the theoretical isochrones from Bertelli et al.~(1994) with
metallicity $Z=0.001$, for the ages 4, 8, 12, 25, 40, and 80 Myr.  On {\it b)}
we also include the 125 Myr isochrone.
On {\it c)} we have plotted the isochrones for 8, 12,
25, 40, and 125 Myr.
On {\it d)} and {\it e)} we have plotted the isochrones, as in {\it c)}, but have
also included the isochrones with ages 60 Myr, and 0.3, 1, and 3 Gyr.  
On each we show the known Cepheids (Piotto et al.~1994) with {\it open circles\/}
surrounding the data points.
A distance modulus of 25.8 mag and an average extinction of $A_V=0.16$ have
been assumed.}
\end{figure}

\clearpage

\begin{figure}
\figurenum{6}
\plotone{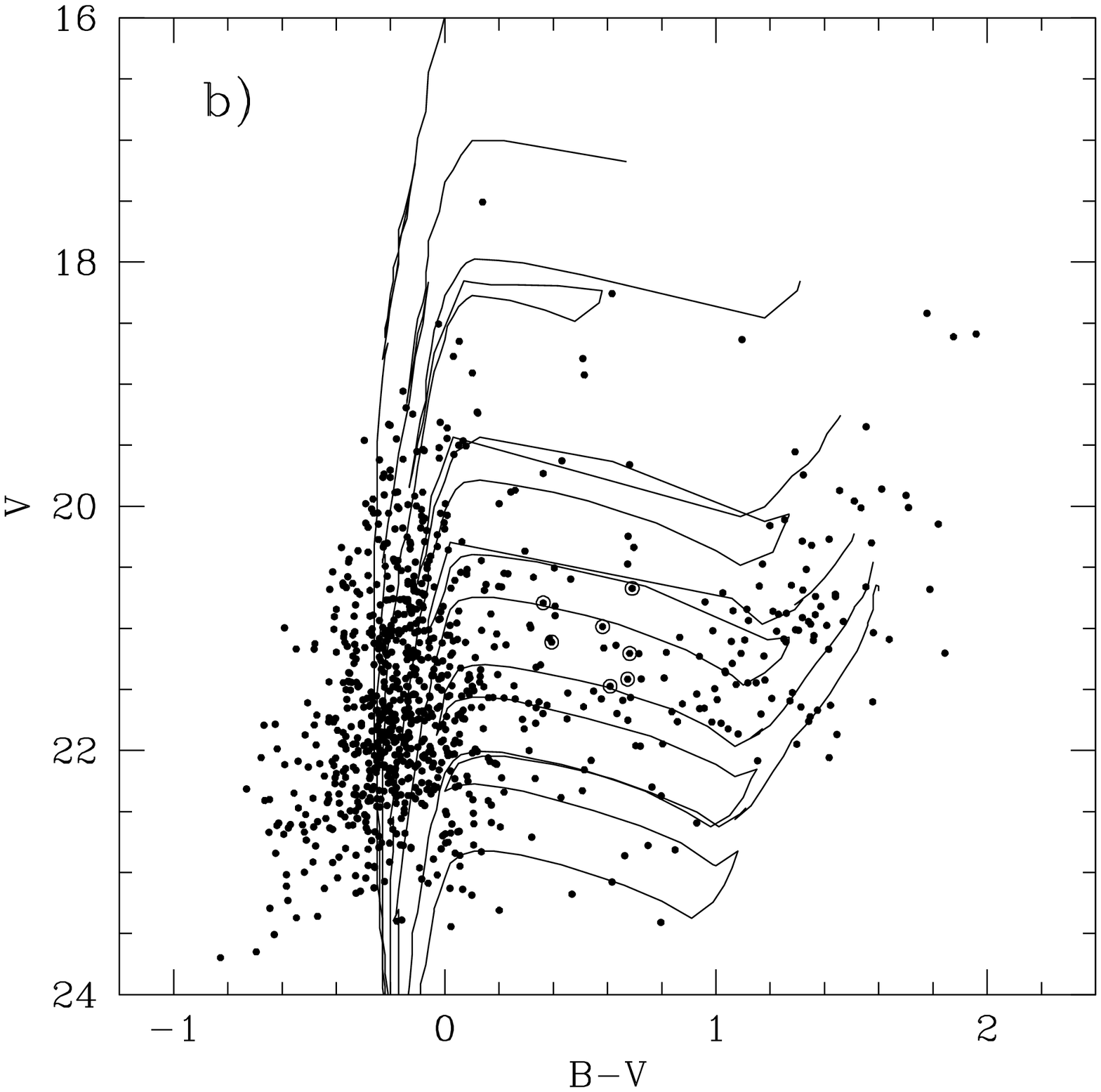}
\caption{(continued)}
\end{figure}

\clearpage

%\begin{figure}
%\figurenum{6}
%\plotone{vandyk.fig6c.ps}
%\caption{(continued)}
%\end{figure}

%\clearpage

%\begin{figure}
%\figurenum{6}
%\plotone{vandyk.fig6d.ps}
%\caption{(continued)}
%\end{figure}

%\clearpage

%\begin{figure}
%\figurenum{6}
%\plotone{vandyk.fig6e.ps}
%\caption{(continued)}
%\end{figure}

%\clearpage

\begin{figure}
\figurenum{7}
%\plotone{vandyk.fig7.ps}
\caption{On the $V$-band image of Sextans A we show
young populations of stars: 1) stars with 
$-1.2 \lesssim U-B \lesssim -1.0$ and $V \lesssim 22$, which are the youngest, bluest
main sequence stars ({\it pluses}); 2) stars with magnitudes and colors, 
particularly in $U-B$, which make them likely main sequence turn-off stars and 
supergiants with ages $\lesssim 12$ Myr ({\it crosses}); 3) blue He-burning stars 
(with $-0.7 \lesssim U-B \lesssim 0.3$) with $V \lesssim 20.7$, i.e., with
ages $\lesssim 50$ Myr ({\it circles}); and, 4) the corresponding red He-burning 
stars, with $1.2 \lesssim V-I \lesssim 1.6$ and $I \lesssim 19.7$ ({\it squares}). }
\end{figure}

%\clearpage

\begin{figure}
\figurenum{8}
%\plotone{vandyk.fig8.ps}
\caption{On the $V$-band image of Sextans A we show
two populations of stars: 1) those blue He-burning stars with ages
$\sim 50$--100 Myr, with $V \gtrsim 20.7$ ({\it squares}); and, 2) the corresponding 
red He-burning stars with $19.7 \gtrsim I \gtrsim 21.3$, with ages to $\sim 100$ Myr
({\it circles}). }
\end{figure}

%\clearpage

\begin{figure}
\figurenum{9}
%\plotone{vandyk.fig9.ps}
\caption{On the $V$-band image of Sextans A we show two older populations: 1) red giants, 
at roughly $I \sim 22$ and $V-I \sim 1.1$, and some AGB stars, with ages between $\sim$100
and $\sim$600 Myr ({\it crosses}); and, 2) those red giants and AGB stars with ages 
$\gtrsim$600 Myr (the older red tangle and red tail stars with ages possibly up to 
$\sim$3 Gyr; [{\it pluses}]). }
\end{figure}

\clearpage

\begin{figure}
\figurenum{10}
%\plotone{vandyk.fig10.ps}
\caption{The distribution of stars in Sextans A with 
$-1.2 \lesssim U-B \lesssim -1.0$ and $V \lesssim 22$, relative to 
the ionized hydrogen (H$\alpha$+[N II] emission). }
\end{figure}

%\clearpage

\begin{figure}
\figurenum{11}
%\plotone{vandyk.fig11.ps}
\caption{The distribution of those stars, as in Figure 10, relative to neutral hydrogen (H I), 
as shown from a column density map derived from 21 cm VLA observations in Graham \& Westpfahl
(1998).  The color bar gives the approximate H~I column density in 
$10^{22}$ cm$^{-2}$.}
\end{figure}

%\clearpage

\begin{figure}
\figurenum{12}
%\plotone{vandyk.fig12.ps}
\caption{The distribution of ionized hydrogen (H$\alpha$+[N II] emission) in Sextans A,
shown in Figure 10, relative to the H I, shown in Figure 11. }
\end{figure}

\clearpage

\begin{figure}
\figurenum{13}
\plotone{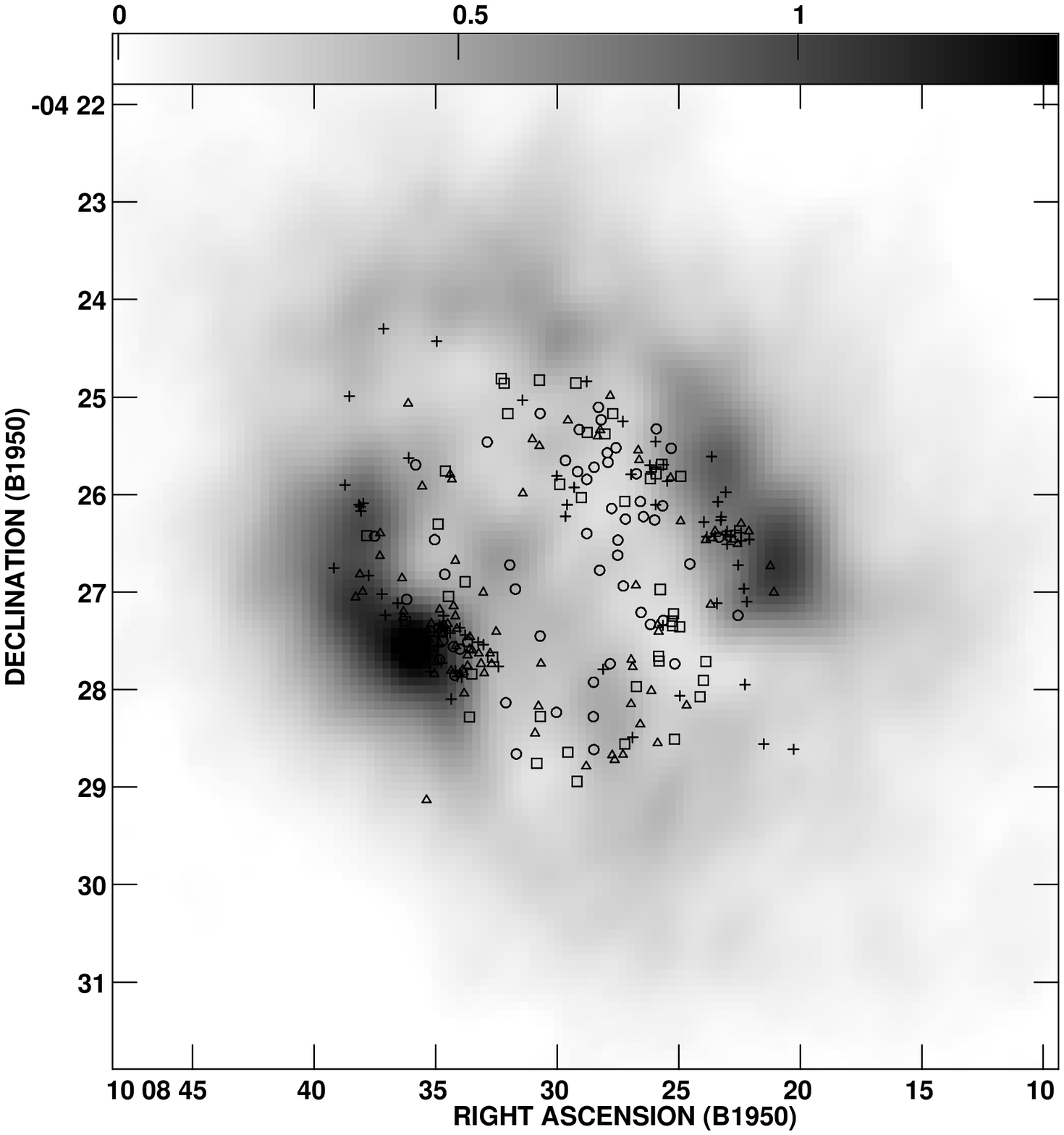}
\caption{The distribution of those stars, as in Figure 7, relative to neutral hydrogen (H I). 
The symbols are the same as in Figure 7, except the likely main sequence turn-off stars and 
supergiants are represented with {\it triangles}, rather than {\it crosses}.  The color bar
is as in Figure 11.}
\end{figure}

\clearpage

\begin{figure}
\figurenum{14}
\plotone{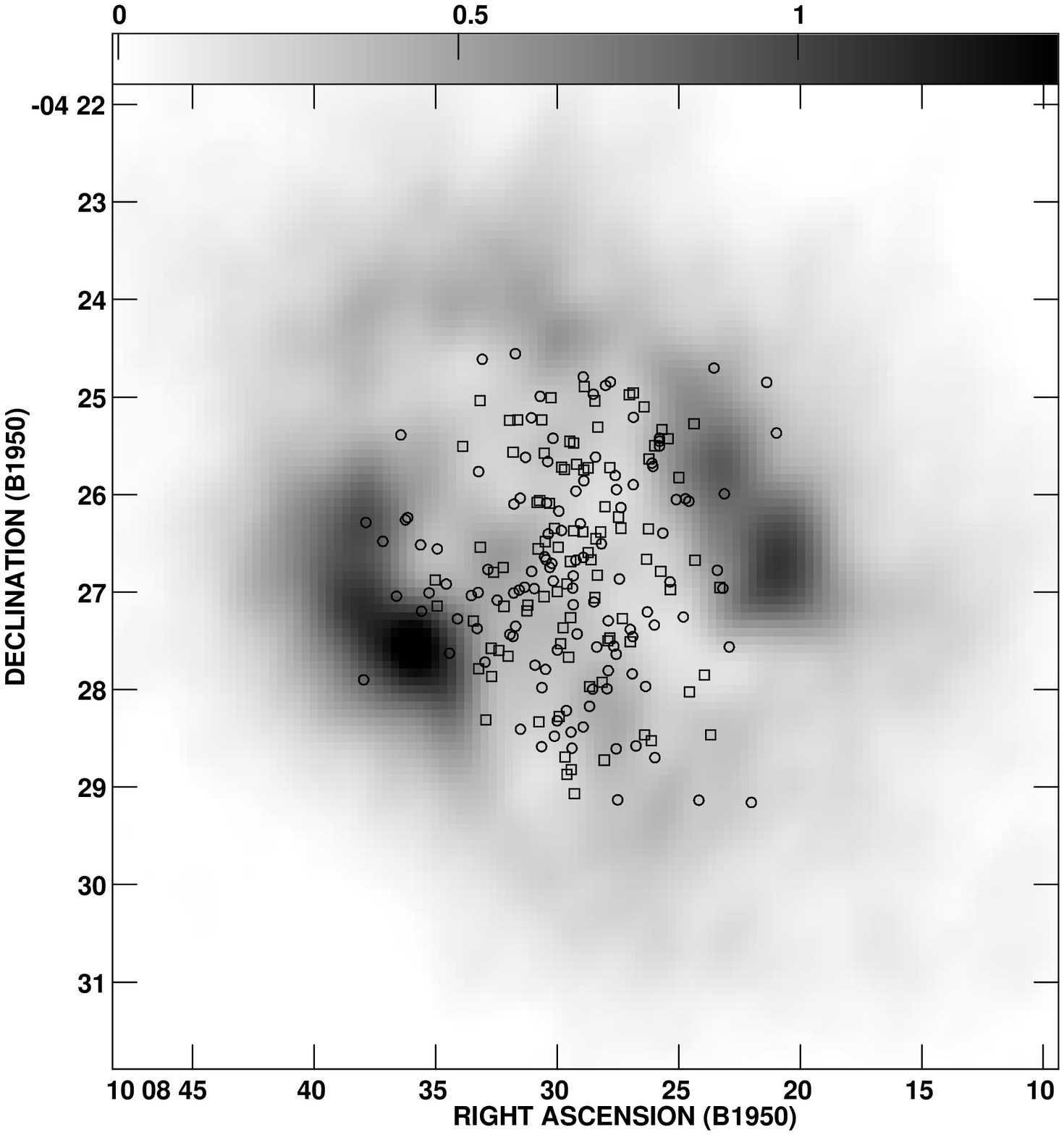}
\caption{The distribution of those stars, as in Figure 8, relative to neutral hydrogen (H~I). 
The symbols are the same as in Figure 8.  The color bar is as in Figure 11.}
\end{figure}

\clearpage

\begin{figure}
\figurenum{15}
\plotone{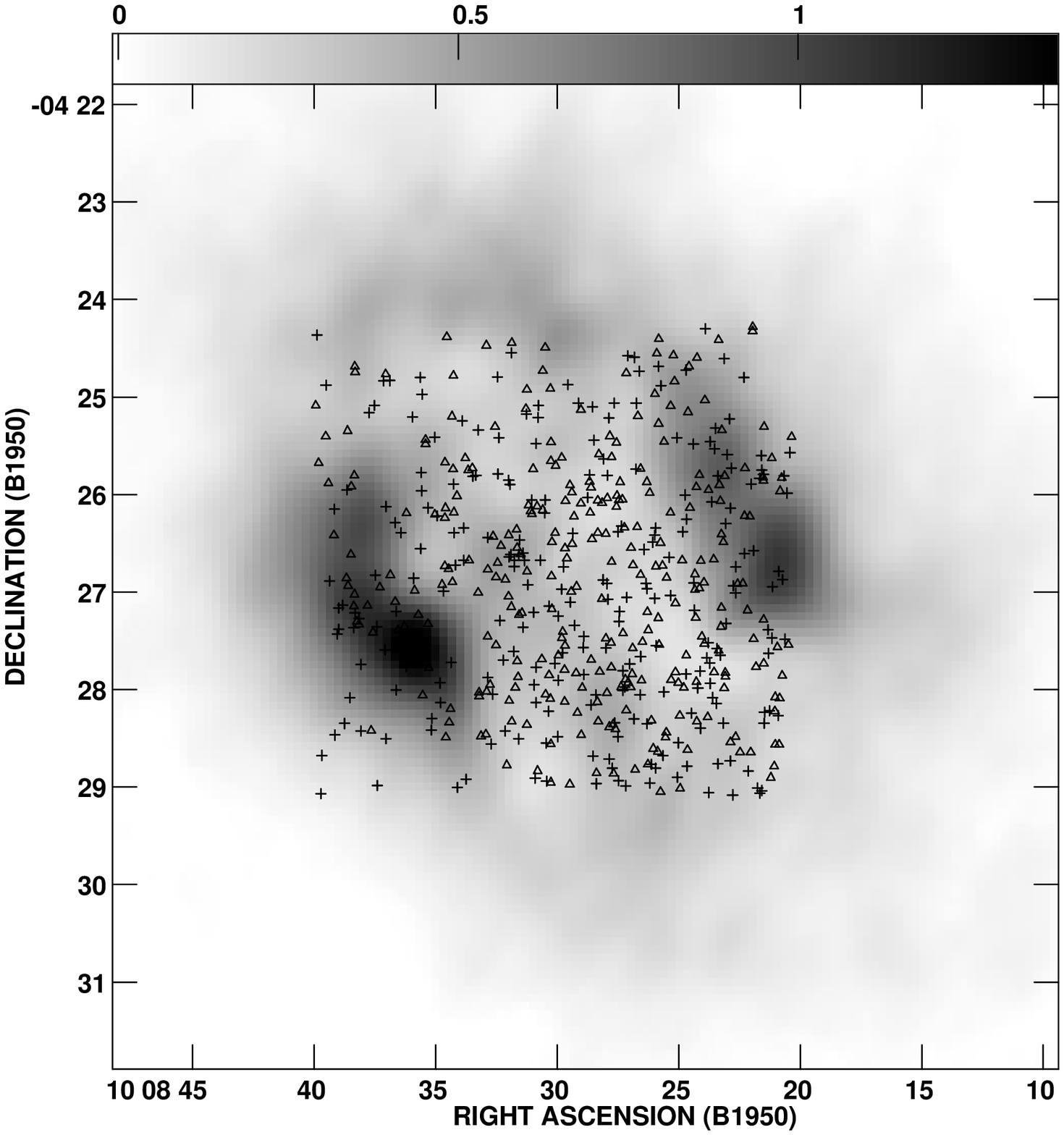}
\caption{The distribution of those stars, as in Figure 9, relative to neutral hydrogen (H~I). 
The symbols are the same as in Figure 9, except the $\sim$100 -- $\sim$600 Myr RGB and AGB stars 
are represented with {\it triangles}, rather than {\it crosses}.  The color bar is as in 
Figure 11.}
\end{figure}

\end{document}